%% file: main_Nature.tex
\documentclass{nature}
\usepackage{graphicx}
\usepackage{amsmath}
\usepackage{amssymb}
\usepackage{color}
\usepackage{url}
\usepackage{verbatim}
\usepackage{tabu}
\usepackage[normalem]{ulem}
\usepackage{sansmath}
\usepackage{upgreek}
\usepackage{epsfig}
\usepackage{caption}
\usepackage[left]{lineno}
\usepackage{setspace}
\usepackage{url}
\usepackage[dvipsnames]{xcolor}
\usepackage{booktabs}

\usepackage[margin=1.1in]{geometry}

\makeatletter
\let\blx@rerun@biber\relax
\makeatother

\DeclareCaptionLabelFormat{adja-page}{\hrulefill\\#1 #2 \emph{(previous page)}}

\DeclareSourcemap{
  \maps[datatype=bibtex]{
    \map[overwrite]{
      \step[fieldsource=month, match=\regexp{\A(j|J)an(uary)?\Z}, replace=1]
      \step[fieldsource=month, match=\regexp{\A(f|F)eb(ruary)?\Z}, replace=2]
      \step[fieldsource=month, match=\regexp{\A(m|M)ar(ch)?\Z}, replace=3]
      \step[fieldsource=month, match=\regexp{\A(a|A)pr(il)?\Z}, replace=4]
      \step[fieldsource=month, match=\regexp{\A(m|M)ay\Z}, replace=5]
      \step[fieldsource=month, match=\regexp{\A(j|J)un(e)?\Z}, replace=6]
      \step[fieldsource=month, match=\regexp{\A(j|J)ul(y)?\Z}, replace=7]
      \step[fieldsource=month, match=\regexp{\A(a|A)ug(ust)?\Z}, replace=8]
      \step[fieldsource=month, match=\regexp{\A(s|S)ep(tember)?\Z}, replace=9]
      \step[fieldsource=month, match=\regexp{\A(o|O)ct(ober)?\Z}, replace=10]
      \step[fieldsource=month, match=\regexp{\A(n|N)ov(ember)?\Z}, replace=11]
      \step[fieldsource=month, match=\regexp{\A(d|D)ec(ember)?\Z}, replace=12]
    }
  }
}

\def\arcmin{$^{\,\prime}$}

\def\lapp{\ifmmode\stackrel{<}{_{\sim}}\else$\stackrel{<}{_{\sim}}$\fi}
\def\gapp{\ifmmode\stackrel{>}{_{\sim}}\else$\stackrel{>}{_{\sim}}$\fi}

\defbibheading{subbibliography}{%
  \section*{}}

\let\cite\autocite

\title{A pulsar-like swing in the polarisation position angle of a nearby fast radio burst}

\input{authors}

\begin{document}

\maketitle

\clearpage
\blfootnote{
\hspace{-0.7cm}
* Corresponding author: Ryan Mckinven. E-mail: ryan.mckinven@mcgill.ca 
\affils
}

\clearpage
\begin{abstract}
Fast radio bursts (FRBs) last for milliseconds and arrive at Earth from cosmological distances. While their origin(s) and emission mechanism(s) are presently unknown, their signals bear similarities with the much less luminous radio emission generated by pulsars within our Galaxy\cite{Pearlman2018} and several lines of evidence point toward neutron star origins\cite{chime/frb2020,Kirsten2021}. For pulsars, the linear polarisation position angle (PA) often exhibits evolution over the pulse phase that is interpreted within a geometric framework known as the rotating vector model (RVM)\cite{Radhakrishnan1969}. Here, we report on a fast radio burst, FRB 20221022A, detected by the Canadian Hydrogen Intensity Mapping Experiment (CHIME) and localized to a nearby host galaxy ($\sim 65\; \rm{Mpc}$), MCG+14-02-011. This one-off FRB displays a $\sim 130$ degree rotation of its PA over its $\sim 2.5\; \rm{ms}$ burst duration, closely resembling the ``S"-shaped PA evolution commonly seen from pulsars and some radio magnetars. The PA evolution disfavours emission models involving shocks far from the source\cite{Lyubarsky2014,Metzger2019,Beloborodov2020} and instead suggests magnetospheric origins\cite{Kumar2017,Zhang2017,Yang2018} for this source which places the emission region close to the FRB central engine, echoing similar conclusions drawn from tempo-polarimetric studies of some repeating sources\cite{Luo2020,Nimmo2022}. This FRB's PA evolution is remarkably well-described by the RVM and, although we cannot determine the inclination and magnetic obliquity due to the unknown period/duty cycle of the source, we can dismiss extremely short-period pulsars (e.g., recycled millisecond pulsars) as potential progenitors. RVM-fitting appears to favour a source occupying a unique position in the period/duty cycle phase space that implies tight opening angles for the beamed emission, significantly reducing burst energy requirements of the source.
\end{abstract}

Like pulsars, the PA evolution of an FRB is an intrinsic property of the source and thus a \textit{direct} probe of the emission mechanism. While FRBs generally display far less evolution in their intraburst PA profiles when compared to pulsars\cite{Pandhi2024}, there exists a subsample of FRBs that do display significant PA variation\cite{Masui2015,Cho2020,Luo2020,Zhang2023}. These observations qualitatively support magnetospheric origins for at least some FRBs, however, the PA variations exhibited by this FRB sample tend to be smaller and more erratic than the smooth, large scale PA evolution commonly encountered in pulsars, suggesting important differences in the emission mechanism of FRBs. 

In the RVM, PA evolution is interpreted as an imprint of the rotation of the neutron star's dipole magnetic field relative to the observer's line-of-sight (LoS). A simple model can be constructed that reproduces the observed PA change as a function of pulse phase, $\phi$,

\begin{equation}
\tan(\rm{PA}-\rm{PA_0})=\frac{\sin \alpha \sin(\phi-\phi_0)}{\sin(\alpha+\beta)\cos(\alpha)-\cos(\alpha+\beta)\sin\alpha\cos(\phi-\phi_0)}.
\label{eqn:rvm}
\end{equation}
Here, $\rm{PA_0}$ and $\phi_0$ represent the position angle and pulse phase of the steepest part of the PA curve while $\alpha$ and $\beta$ are parameters characterizing the geometry of the system. In particular, $\alpha$ represents the inclination angle between the neutron star's magnetic axis and rotation axis and $\beta$ is the impact angle between the LoS and the magnetic axis. While some pulsars display complex polarimetric behaviour (e.g., orthogonal modes\cite{Stinebring1984,McKinnon2000}), many pulsars exhibit smooth ``S"-shaped swings in their PA that are a natural prediction of the RVM and serve as an important indicator supporting radio emission occurring near the poles of a rotating dipolar magnetosphere.


FRB 20221022A was detected on October 22, 2022 by CHIME/FRB\cite{chime/frb2018}. The burst was sufficiently bright to trigger the readout of the raw voltage (baseband) data from each of CHIME's 1024 dual-polarised feeds through the instrument's baseband backend. Subsequent processing with CHIME/FRB's baseband pipeline\cite{Michilli2021} provided a refined localization in equatorial coordinates of
$\rm{RA, Dec = 03\,h\,14\,m\,31\,s(22), +86^{\circ}52'19''(14)}$\footnote{$1\sigma$ uncertainties} (J2000), greatly improving but consistent with the initial localization that was circulated to the wider FRB community through the CHIME/FRB Virtual Observatory Event (VOEvent) Service\footnote{https://www.chime-frb.ca/voevents}.

\begin{figure}
\begin{center}
\includegraphics[width=0.44\textwidth]
{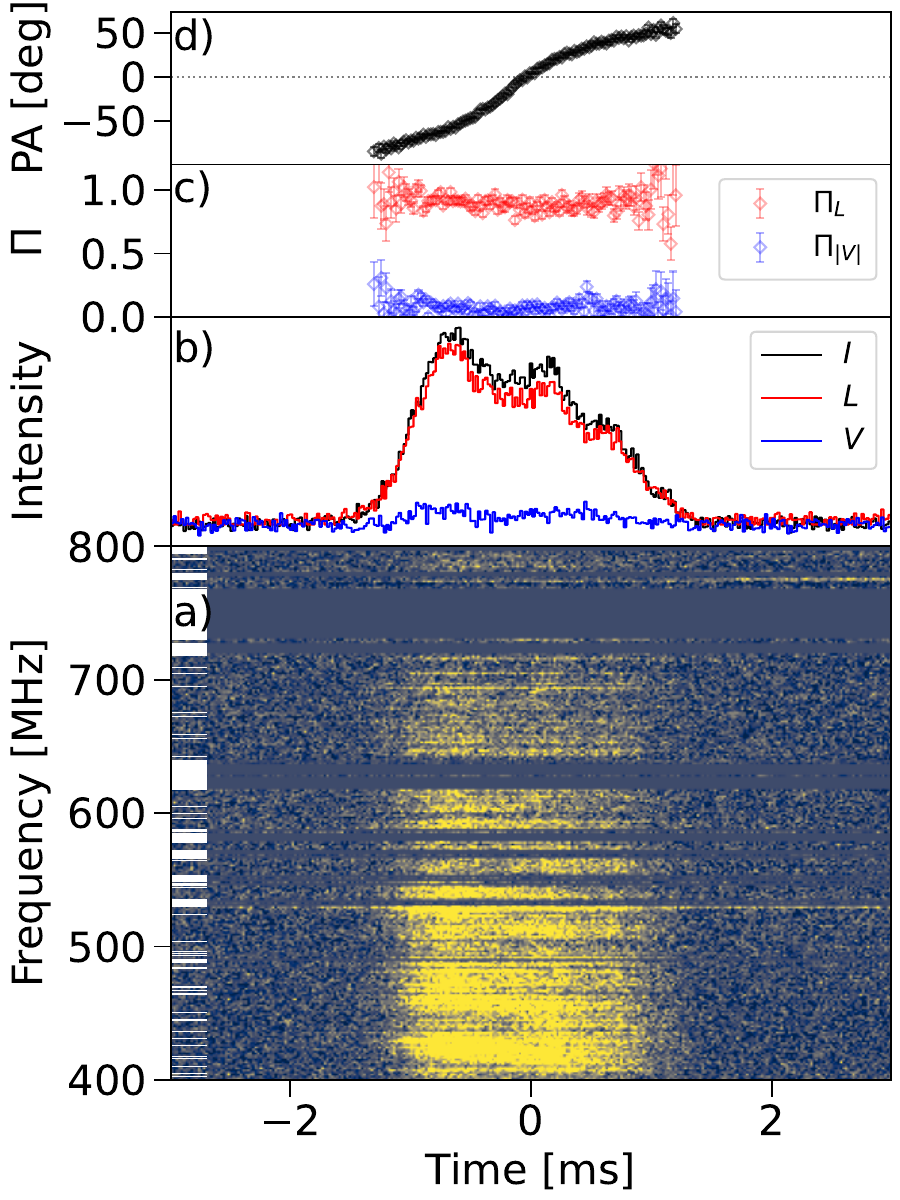}
\includegraphics[width=0.47\textwidth]
{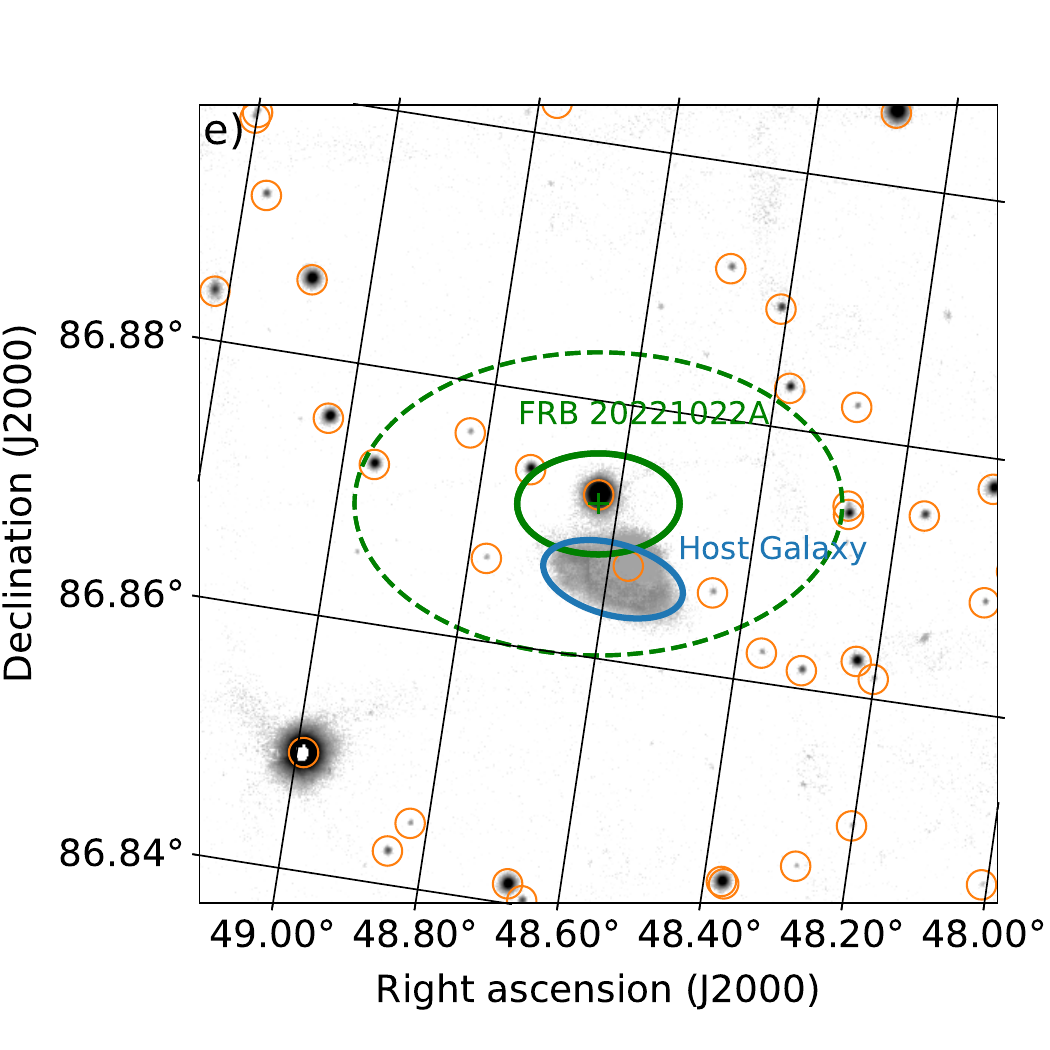} \\
\caption{a) Total intensity waterfall (time-frequency) plot of FRB 20221022A produced from baseband data and coherently dedispersed to the structure optimizing DM ($\rm{DM_{struct}}$; see Table~\ref{tab1}). Data have been rebinned in both time and frequency at resolutions $20.48\;\rm{\mu s}$ and 1.5625 MHz, respectively. Masked or missing frequency channels are indicated by white lines along the left vertical axis. b) Burst profiles of total (black), linear polarised (red), and circularly polarised (blue) intensity that are obtained by integrating the signal over the entire $400-800$ MHz CHIME band. c) Measurements of $\Pi_L$ and $\Pi_V$ over the burst duration. d) Polarisation position angle (PA) measurements demonstrating the ``S"-like evolution over the burst duration. e) Pan-STARRS 3$\pi$ (PS1) $r$-band image\cite{Waters16} of the approximate localization region of FRB 20221022A, represented by $1\sigma$ (solid) \& $3\sigma$ (dashed) green error ellipses.
Foreground stars are indicated by orange markers taken from the Gaia catalog (DR3)\cite{Gaia2023}. The host galaxy, MCG+14-02-011, is encircled by a blue ellipse (encompassing $95\%$ of the flux), corrected for contaminant foreground star light (see Methods and Extended Data Fig.~\ref{fig:cutout_unmasked} for the uncorrected image).}
\label{fig:waterfall}
\end{center}
\end{figure}  

Baseband data beamformed on the best localization of FRB 20221022A were used to construct Stokes $I,Q,U,V$ polarisation parameters. Fig.~\ref{fig:waterfall} displays the radio signal of FRB 20221022A, corrected for the frequency-dependent phase/delay of the signal's transit through the dispersive intervening medium using a structure-maximizing dispersion measure, $\rm{DM}=116.8371(14)\; \rm{pc\, cm^{-3}}$, that effectively aligns the burst so that temporal substructure is simultaneous with respect to frequency\cite{Seymour2019} (see Methods). At this DM, the burst is optimally decomposed into three partially overlapping subcomponents, all of which are broadband, with the emission stronger towards the bottom of the band (see Methods). Temporal scattering from multi-path propagation is modest for this FRB and is difficult to directly measure due to the absence of prominent microstructure. An upper limit on the temporal scattering is thus conservatively determined to be $\lesssim550\; \rm{\mu s}$, corresponding to twice the shortest duration sub-burst.
Assuming a ``thin-screen" formalism for scatter-broadening that predicts $\tau \propto \nu^{-4}$, we scale this measurement to 1 GHz and obtain an upper limit on scattering $\tau_{\rm{1 GHz}}\lesssim 70\; \mu s$ that is consistent with predictions from Galactic free electron density models\cite{ne2001,ymw17} (Table~\ref{tab1}). The degree of fractional linear polarisation ($\Pi_L$) of the burst is high, $\Pi_L\gtrsim 95\%$, with no evident time dependence. We check for frequency dependence in  $\Pi_L$ by measuring the fractional linear polarisation over the 400-600 MHz and 600-800 MHz subbands and retrieve values that are consistent with each other and the band-averaged value, indicating very little if any depolarisation. Taking the ratio of $\Pi_L$ over the two subbands provides a value, $f_{\mathrm{depol}} = \Pi^{600-800}_L / \Pi^{400-600}_L=0.95(8)$, that quantifies the absence of evidence for depolarisation in a manner that does not depend on the assumed depolarisation model\cite{Feng2022,Beniamini2022} and is consistent with equivalent measurements drawn from a subsample of broadband one-off FRBs recently reported from CHIME/FRB\cite{Pandhi2024}. Meanwhile, the small amount of circular polarisation does not exceed $10\%$ and can easily be instrumental based on our understanding of residual systematics of CHIME\cite{Mckinven2021} (see Methods). A Faraday rotation measure (RM) of $\rm{RM}=-40.39_{-0.02}^{+0.03}\; \rm{rad\, m^{-2}}$ was found, which after accounting for the Galactic RM contribution (see Table~\ref{tab1}), implies a modest contribution of $\sim -30\; \rm{rad\, m^{-2}}$ from the host galaxy and the source's local environment. The measured $\rm{RM}$ was used to determine the PA curve by derotating the modulation in Stokes $Q,U$ and integrating the signal over the 400-800 MHz band (see Methods). 

The resulting ``S"-shaped swing in the PA displayed in panel (d) of Fig.~\ref{fig:waterfall} is remarkably pulsar-like and is the first of its kind for a one-off FRB, with a total PA excursion that is substantially larger than that seen from the diverse PA curves reported from repeating source FRB 20180301A\cite{Luo2020}. Given the low DM of FRB 20221022A and the modest offset from the Galactic plane ($\ell\sim 24.6$ degrees), we checked that the source is indeed an FRB and not a misclassified or unidentified Galactic source. Cross-matching the source localization ellipse with existing galaxy catalogs led to the identification of the galaxy MCG+14-02-011 as the probable host (panel (e) of Fig.~\ref{fig:waterfall}). This association was confirmed after careful subtraction of contaminant light from a bright foreground star, which provided an improved estimate on the host galaxy r-band brightness of $15.1(3)$ (AB mag) and a probability of association of $\gtrsim 99 \%$ as determined using the Probabilistic Association of Transients to their Hosts (PATH) methodology\cite{Aggarwal2021} (see Methods). Our confidence in MCG+14-02-011 as the host was independently bolstered by spectroscopic followup with the Gran Telescopio Canarias (GTC) that established the galaxy as nearby with a redshift of 0.0149(3), consistent with the low excess (extragalactic) dispersion measure, $\rm{DM_{extra}}$, of this event after subtracting an estimate of the Galactic DM contribution (see Table~\ref{tab1}). We detect significant frequency modulation in the burst spectrum that is consistent with scintillation from multi-path interference. We determine two scintillation scales in the burst spectrum indicating the presence of two scattering screens, one residing in our own Galaxy and another likely located in the host galaxy, further supporting the extragalactic nature of FRB~20221022A despite its very low DM excess (see Methods; Nimmo et al. in preparation). 
Spectroscopic analysis of GTC data identified several emission lines (e.g., [O$_{\rm{III}}$]) that indicate the presence of star formation, supporting MCG+14-02-011 as a late type galaxy (see Methods). 
 
\begin{figure}[h]%
\centering
\includegraphics[width=0.99\textwidth]{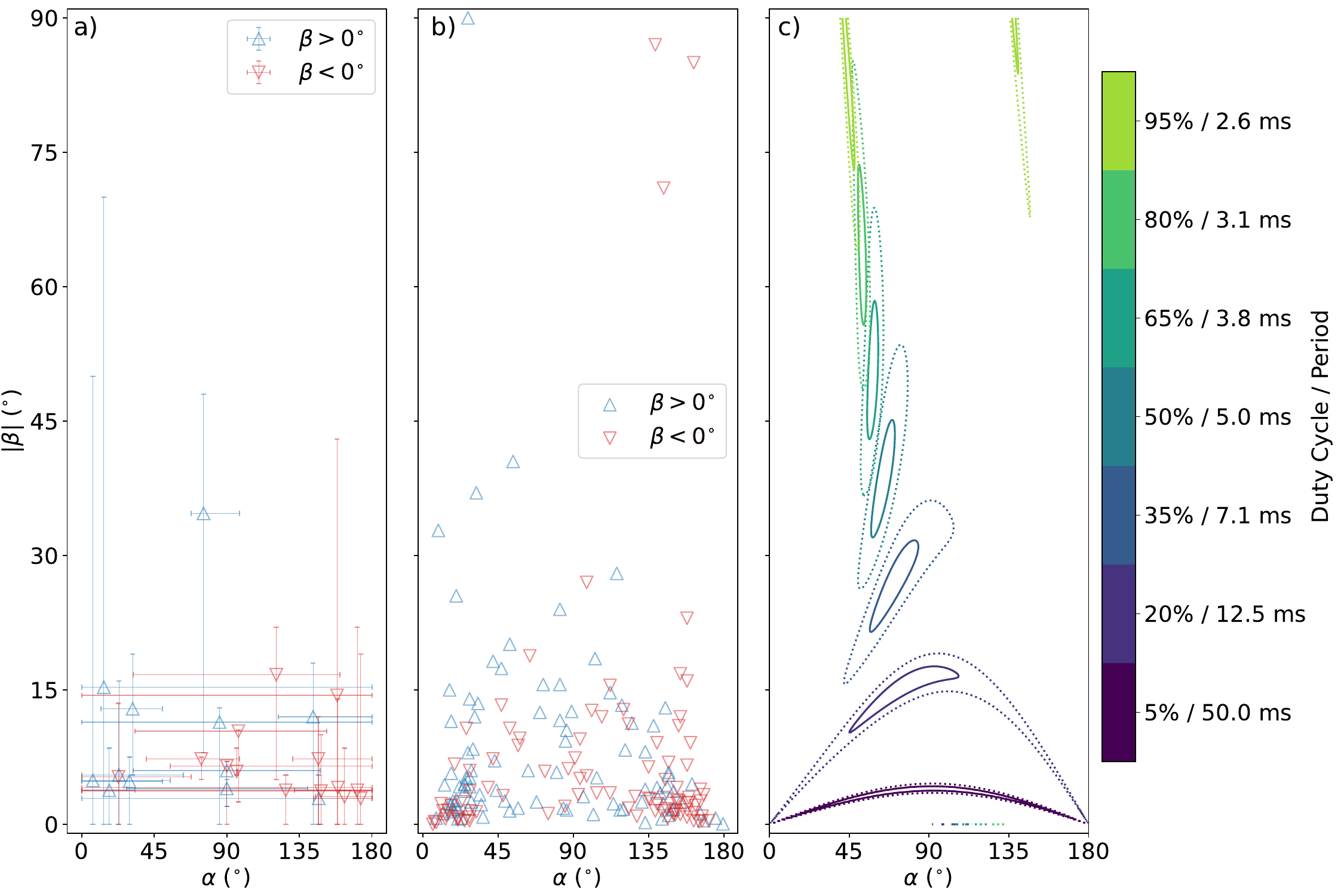}
\caption{Constraints on RVM $\alpha, \beta$ parameter phase space for a) a pulsar sample observed by Parkes (Murriyang) radio telescope\cite{Rookyard2015}, b) a pulsar sample observed by the Five-hundred-meter Aperture Spherical Telescope (FAST)\cite{Wang2023} and c) FRB 20221022A studied here for different assumed values of the duty cycle/period of the source. The contours of panel (c) correspond to $1\sigma$ (solid line) \& $3\sigma$ (dotted line) uncertainties and the quality of the model fits are observed to worsen at higher assumed duty cycles (see Extended Data Fig.~\ref{fig:PA_curve_fits}). Measurement uncertainties have been omitted in panel (b) for clarity.}
\label{fig:alpha_beta}
\end{figure}

Having established FRB 20221022A as extragalactic from a robust association to its host galaxy, we investigate its striking PA evolution. Analysis of a large sample of one-off (apparently non-repeating) FRBs detected by CHIME/FRB indicates significant PA variations in approximately $10\%$ of such sources\cite{Pandhi2024}. The PA evolution in this sample tends to be characterized by modest variations interspersed by erratic PA jumps across burst sub-components. Relative to this sample, the large PA swing reported here from FRB 20221022A is anomalous (see Extended Data Fig.~\ref{fig:PA_curve_stats}) and the smooth and sustained PA evolution suggests a geometric effect linking the change in PA with the burst duration. Beamed emission sweeping through our LoS would naturally explain this behaviour, with the symmetry of FRB 20221022A's PA swing indicating a highly ordered magnetic field near the source. Conditions for such behaviour are naturally found in the magnetospheres of neutron stars, known to preferentially beam radio emission along magnetic poles that, to first order, are well-described as dipolar\cite{Philippov2022}. This interpretation is supported by the gradient of the PA swing, $d\rm{\rm{PA}}/dt$, being maximal near the centroid of the burst envelope (see Fig.~\ref{fig:PA_curve_fits}), a common occurrence in pulsar PA curves and a prediction of core/conal pulsar emission models that has been leveraged to measure the altitude of the emission from the neutron star surface via the Aberration/Retardation (A/R) effect\cite{Blaskiewicz1991}. Such properties of the PA curve of FRB 20221022A are thus difficult to explain as a random coincidence not tied to an underlying geometric effect of an ordered magnetic field. This interpretation is strongly supported by the remarkable symmetry of the PA swing both with respect to the burst centroid and the PA inflection point; for such behaviour to occur by chance would seem unlikely given the PA behaviour that is typically seen from FRBs. Beamed emission from a rotating neutron star is not the only scenario and it remains plausible that analogous conditions for beamed emission can be produced from other compact sources. Launched jets from accreting massive back holes have been shown to display PA variations\cite{Blinov2015,Blinov2016}, however, these variations appear over much longer timescales of minutes to hours and are possibly an imprint of the jet's precession\cite{Miller-Jones2019} or some other geometric effect\cite{Lyutikov2017}. While these accreting black hole models are interesting to consider, the timescale would necessitate a stellar-mass sized black holes, which are not well-studied objects with regards to polarimetry.

Given the strong evidence for neutron star origins of FRBs, we thus focus our investigation on extending the RVM (Equation~\ref{eqn:rvm}) to fit the PA swing of FRB 20221022A. This assumes an underlying periodicity, unmeasurable here given just one burst, which makes the RVM underdetermined. However, by making assumptions on the period and, by extension, on the duty cycle of the source (i.e., the fraction of the period where emission is observed), we find that the observed PA evolution is well-replicated by the RVM. In order to better understand how the assumed period of the source affects the resulting best-fit geometrical parameters of the RVM, we perform a grid search over $\alpha,\beta$ space at trial duty cycles of 5\%, 20\%, 35\%, 50\%, 65\%, 80\%, 95\% (see Methods). Panel (c) of Fig.~\ref{fig:alpha_beta} shows the resulting likelihood contours for different assumed duty cycles/periods of the source. Equivalent constraints on $\alpha,\beta$ for a sample of pulsars\cite{Rookyard2015,Wang2023} are shown in panels (a) and (b) of Fig.~\ref{fig:alpha_beta}, where we see a tendency for clustering near $\alpha=0$ or 180 degrees, indicating a general preference for alignment of the magnetic and rotation axis of the neutron star. Meanwhile, constraints on $|\beta|$ values of pulsars indicate a preference for modest values, a reflection of the greater chance of detecting beamed pulsar emission when the angle subtended by the LoS and the neutron star's magnetic dipole is small. 

The likelihood contours of panel (c) are strongly dependent on the assumed period/duty cycle of FRB 20221022A. Associated best-fit PA curves for each trial duty cycle/period (see Extended Data Fig.~\ref{fig:PA_curve_fits}) indicate a worsening fit quality at duty cycles $\gtrsim 50\%$ ($\rm{period\lesssim 5\; ms}$), effectively ruling out extremely short period pulsars (e.g. recycled millisecond pulsars) as the source. Meanwhile, in the long period regime ($\rm{period\gtrsim 5\; ms}$), the fit quality of the RVM remains stable, clustering at reduced $\chi^{2}\lesssim 2$ and effectively insensitive to the specific duty cycle/period. We confirm this by performing a second grid search over $\alpha, \beta$ for a sample of extremely short duty cycles ($2.5,1.0,0.5,0.1\%$) and retrieve similar fit quality with constraints on $\alpha$ that effectively span the entire $0-180$ degrees range (see Methods). In this long period regime, the combination of small duty cycle and $\beta$ values imply tight opening angles for the beamed emission within the RVM framework (see Extended Data Fig.~\ref{fig:alpha_beta_vs_rho} \& \ref{fig:alpha_beta_vs_rho2}). Such highly beamed emission would greatly reduce the total burst energy compared to isotropic emission. 

\begin{table}
\caption{Properties associated with FRB 20221022A}\label{tab1}%
\small
\begin{center}
\scalebox{0.78}{
\begin{tabular}{@{}ll@{}}
\toprule
\textbf{Parameter} & \textbf{Value$^a$} \\
\midrule
\multicolumn{2}{c}{General}  \\
\hline
Right Ascension$^b$ (RA; J2000) & $\rm{03\, h\, 14\, m\, 31\, s(22)}$  \\
Declination$^b$ (Dec; J2000) & +86$^{\circ}$52$'$19$''$(14) \\
CHIME arrival time (at 400 MHz; UTC) & 2022-10-22 10:14:59 \\ 
$\rm{W_{10\%}}^c$ & $2.5(2)\; \rm{ms}$ \\ 
S/N (box-car) & 284.7 \\
Flux$^d$ & $55(6)\; \rm{Jy}$ \\
Fluence$^d$ & $87(11)\; \rm{Jy\, ms}$ \\
$\rm{E}^d$ (isotropic) & $2.62(33)\times 10^{38}\; \rm{erg}$ \\
$\rm{\dot E}^d$ (isotropic) & $1.05(16)\times 10^{41}\; \rm{erg\, s^{-1}}$ \\
Dispersion Measure, DM$^e$ & $116.8371(14)\; \rm{pc\, cm^{-3}}$ \\
$\rm{DM}_{\rm{MW-NE2001}}^f$ & $60(12)\; \rm{pc\,cm^{-3}}$ \\
$\rm{DM}_{\rm{MW-YMW16}}^g$ & $60(12)\; \rm{pc\,cm^{-3}}$ \\
$\rm{DM}_{\rm{MW-halo}}$ & $30(20)\; \rm{pc\, cm^{-3}}$ \\
$\rm{DM}_{\rm{extragalactic}}$ & $27(23)\; \rm{pc\, cm^{-3}}$ \\
$\rm{DM}_{\rm{IGM}}$ &  $\gtrsim 13\; \rm{pc\, cm^{-3}}$ \\
$\tau_{\rm{1GHz-NE2001}}^f$ & $0.41\,\rm{\mu s}$ \\
$\tau_{\rm{1GHz-YMW16}}^g$ & $2.7\,\rm{\mu s}$ \\
Exposure & 1090 hrs (upper) / 1662 hrs (lower) \\
\hline
\multicolumn{2}{c}{Host Galaxy}  \\
\hline
redshift, $z$ & $0.0149(3)$\\
$\rm{D_L}^h$ & $65.2(1.3)\; \rm{Mpc}$ \\
$\rm{DM}_{\rm{host}}/(1+z)$ & $\lesssim 14(23)\; \rm{pc\, cm^{-3}}$ \\
Present-day Stellar Mass, $\log (M^{\star}/M_{\odot})$ & $\sim 9.6$\\
Total Star Formation Rate (SFR) & $\sim$ 0.47 $\rm M_{\odot} yr^{-1}$ \\
\hline
\multicolumn{2}{c}{Polarisation}  \\
\hline
$\text{RM}$ & $-40.39_{-0.02}^{+0.03}\; \rm{rad\, m^{-2}}$ \\
$\text{RM}_{\text{MW}}^i$ & $-8.9(2.5) \; \rm{rad\, m^{-2}}$ \\
$\langle \Pi_{\rm{L}} \rangle$ & $95(3)\%$ \\
$\langle \Pi_{\rm{V}} \rangle$ & $\lesssim 10\%$\\
$\Pi^{\rm{400-600 MHz}}_{\text{L}}$ & $96(4)\%$ \\
$\Pi^{\rm{600-800 MHz}}_{\text{L}}$ & $91(7)\%$ \\
\hline
\multicolumn{2}{c}{Fitburst} \\
\hline
S/N$^j$ & 267.9 \\
DM & $116.843(1)\; \rm{pc\, cm^{-3}}$ \\
\# of components & 3 \\
Width$^k$ & $298.7^{+8.7}_{-9.0}\; \rm{\mu s}, 282^{+23}_{-26}\; \rm{\mu s}, 277^{+25}_{-22}\; \rm{\mu s}$\\
$\gamma^k$ & $0.96(20)$, $-1.01^{+0.25}_{-0.23}$, $-0.43^{+0.47}_{-0.39}$\\
$r^k$ & $-8.6(5)$, $-5.5(6)$, $-4.9^{+0.9}_{-0.8}$ \\
\toprule
\end{tabular}
}
\end{center} 
$^a$ Uncertainties are reported at the $1\sigma$ confidence level \\
$^b$ Seconds of arc as distances on the sky \\
$^c$ Burst width at 10\% of peak (observer's frame) \\
$^d$ Calculated for the entire 400-800 MHz CHIME band \\
$^e$ Structure-maximizing DM determined with \texttt{DM\_phase}\cite{Seymour2019} \\
$^f$ Galactic DM and scattering estimates determined from the NE2001 electron density model\cite{Cordes2002,Cordes2003} \\
$^g$ Galactic DM and scattering estimates determined from the YMW16 electron density model\cite{Yao2017} \\
$^h$ Luminosity distance, assuming a (WMAP 9-year)\cite{Hinshaw2013} Hubble constant of $H_0=69.32\; \rm{km\, Mpc^{-1}\, s^{-1}}$ \\
$^i$ Galactic RM estimate as determined from a Galactic Faraday rotation sky map\cite{Hutschenreuter2022} \\
$^j$ S/N determined here is significantly larger than the real-time detection due to enhanced sensitivity of beamformed baseband data \\
$^k$ Measurements of width, spectral index ($\gamma$) and spectral running ($r$) are determined separately for each of the three burst subcomponents
\end{table}

The isotropic burst luminosity of FRB 20221022A is $\rm{\dot E_{\rm{iso}}} = 1.05(16)\times 10^{41}\; \rm{erg\, s^{-1}}$ (see Methods). The comparison of this value to Galactic pulsar spin-down luminosities reveals a substantial disparity that is further exacerbated by the inclusion of the radio efficiency factor, $\xi$. This factor accounts for the total spin-down power that is emitted as radio waves\cite{Szary2014} and is typically $\xi \sim 10^{-8}-10^{-6}$ for pulsars with spin-down luminosities $10^{36}-10^{37}$ erg s$^{-1}$ and $\xi \sim 10^{-3}-10^{-1}$ if the spin-down luminosity is $< 10^{31}$ erg s$^{-1}$. While this disparity in the energy budget can be partially mitigated if the emission of FRB 20221022A is preferentially beamed, the required opening angles ($\ll$1 degree) would be significantly smaller than that commonly measured for pulsars and imply a duty cycle/period combination that is distinct relative to the observed Galactic pulsar sample (see Extended Data Fig.~\ref{fig:pulsar_compare}). The absence of similar pulsar measurements in this phase space challenges the notion of a small beaming angle, suggesting a possible undetected population of long-period, transient radio-emitting neutron stars. This idea finds support in the characteristics of young pulsars with large spin-down luminosities, resembling FRB 20221022A, showing substantial linear polarisation (see Methods). Indeed, $\Pi_L$ measurements are known to correlate with inferred spin-down luminosities for an ATNF-catalog pulsar sample\cite{Johnston2023,Wang2023}, although the origins of this correlation are not understood. If FRB 20221022A originates from a young pulsar, its high polarisation aligns with expectations for a source with an anomalously large energy loss rate. Alternative interpretations exist however, as seen in radio magnetars with high $\Pi_L$ and distinctive position angle swings. Regardless of the uncertainty on the energy source and emission mechanism, the PA swing of FRB 20221022A appears to be well-described by a simple geometric RVM model. The remarkably large and smooth PA swing is anomalous for the one-off FRB sample and suggests either a geometric selection effect that deters detection of such sources or an altogether different class of source and/or emission mode.

\printbibliography[segment=\therefsegment,heading=subbibliography]

\clearpage

\clearpage
\input{methods}

\printbibliography[segment=\therefsegment,heading=subbibliography,filter=notother]

\begin{addendum}
\item[Data availability]
The beamformed baseband data needed to reconstruct the PA measurements of this source are stored within a Hierarchical Data Format 5 file and is available at \url{<URL>}. The GTC data are available upon request.

\item[Code availability]
The code used to the perform the grid search of the RVM parameter space is available at \url{<URL>}.

\item
\input{acknowledgements}
\genacks
\allacks
    \item[Author Contributions]
    R.M. led the analysis and interpretation of the baseband data recorded by CHIME/FRB, including the polarisation analysis and RVM-fitting, in addition to the paper writing. 
    M.B. and A.K. conducted the GTC observations, including the data reductions, spectroscopic analysis and interpretation.
    A.B.P. and M.B. searched for multiwavelength counterparts in archival data and wrote the corresponding sections of the paper.
    T.E. and C.P. contributed to analysis supporting the Pcc estimate of the putative host galaxy via PATH and analysis of Pan-STARRS data.
    A.P. contributed to the development and preliminary application of the RVM-fitting code.
    A.C. worked on improved estimates of the Galactic DM contribution and, along with Z.P., contributed to the repetition search.
    U.G. and K.S. contributed to the burst modelling via fitburst.
    D.M. obtained the flux/fluence measurements and K.N. led the scintillation analysis.
    All authors contributed to the discussion of the results presented and commented on the manuscript.
    
    \textbf{Correspondence and requests for
        materials} should be addressed to Ryan Mckinven\\(email: ryan.mckinven@mcgill.ca).
        
\end{addendum}


\end{document}

%% file: authors.tex
\author{
Ryan Mckinven$^{1,2}$, \allowbreak
Mohit Bhardwaj$^{3}$, \allowbreak
Tarraneh Eftekhari$^{6,22}$, \allowbreak
Charles D. Kilpatrick$^{6}$, \allowbreak
Aida Kirichenko$^{9,31}$, \allowbreak
Arpan Pal$^{12}$, \allowbreak
Amanda M. Cook$^{4,5}$, \allowbreak
B. M. Gaensler$^{4,5,7}$, \allowbreak
Utkarsh Giri$^{8}$, \allowbreak
Victoria M.~Kaspi$^{1,2}$, \allowbreak
Daniele Michilli$^{10,11}$, \allowbreak
Kenzie Nimmo$^{10,11}$, \allowbreak
Aaron B. Pearlman$^{1,2}$, \allowbreak
Ziggy Pleunis$^{4,5,33,34}$, \allowbreak
Ketan R. Sand$^{1,2}$, \allowbreak
Ingrid Stairs$^{13}$, \allowbreak
Bridget C. Andersen$^{1,2}$, \allowbreak
Shion Andrew$^{10,11}$, \allowbreak 
Kevin Bandura$^{14,15}$, \allowbreak
Charanjot Brar$^{1,2}$ \allowbreak
Tomas Cassanelli$^{16}$, \allowbreak
Shami Chatterjee$^{17}$, \allowbreak
Alice P. Curtin, $^{1,2}$, \allowbreak
Fengqiu Adam Dong$^{13}$, \allowbreak
Gwendolyn Eadie$^{5,18,32}$, \allowbreak
Emmanuel Fonseca$^{15,19}$, \allowbreak
Adaeze L. Ibik$^{4,5}$, \allowbreak
Jane F. Kaczmarek$^{20}$, \allowbreak
Bikash Kharel$^{15,19}$, \allowbreak
Mattias	Lazda$^{4,5}$, \allowbreak
Calvin Leung$^{21,22}$, \allowbreak
Dongzi Li$^{23}$, \allowbreak
Robert Main$^{1,2}$, \allowbreak
Kiyoshi W. Masui$^{10,11}$, \allowbreak
Juan Mena-Parra$^{4,5}$, \allowbreak
Cherry Ng$^{27}$, \allowbreak
Ayush Pandhi$^{4,5}$, \allowbreak
Swarali Shivraj Patil$^{15,19}$, \allowbreak
J. Xavier Prochaska$^{24,25,26}$, \allowbreak
Masoud Rafiei-Ravandi$^{1,2}$, \allowbreak 
Paul Scholz$^{4,28}$, \allowbreak  
Vishwangi Shah$^{1,2}$, \allowbreak
Kaitlyn Shin$^{10,11}$, \allowbreak 
Kendrick Smith$^{29}$ \allowbreak
}
\newcommand{\affils}{
\begin{affiliations}
\item{Department of Physics, McGill University, 3600 rue University, Montr\'eal, QC H3A 2T8, Canada}
\item{Trottier Space Institute, McGill University, 3550 rue University, Montr\'eal, QC H3A 2A7, Canada}
\item{McWilliams Center for Cosmology, Department of Physics, Carnegie Mellon University, Pittsburgh, PA 15213, USA}
\item{Dunlap Institute for Astronomy \& Astrophysics, University of Toronto, 50 St.~George Street, Toronto, ON M5S 3H4, Canada}
\item{David A.~Dunlap Department of Astronomy \& Astrophysics, University of Toronto, 50 St.~George Street, Toronto, ON M5S 3H4, Canada}
\item{Center for Interdisciplinary Exploration and Research in Astrophysics (CIERA) and Department of Physics and Astronomy, Northwestern University, Evanston, IL 60208, USA}
\item{Division of Physical and Biological Sciences, University of California Santa Cruz, Santa Cruz, CA 95064, USA}
\item{Department of Physics, University of Wisconsin-Madison, 1150 University Ave, Madison, WI 53706, USA}
\item{Instituto de Astronom\'ia, Universidad Nacional Aut\'onoma de M\'exico, Apdo. Postal 877, Ensenada, Baja California 22800, M\'exico}
\item{MIT Kavli Institute for Astrophysics and Space Research, Massachusetts Institute of Technology, 77 Massachusetts Ave, Cambridge, MA 02139, USA}
\item{Department of Physics, Massachusetts Institute of Technology, 77 Massachusetts Ave, Cambridge, MA 02139, USA}
\item{National Centre for Radio Astrophysics, Tata Institute of Fundamental Research, Pune 411007, India}
\item{Department of Physics and Astronomy, University of British Columbia, 6224 Agricultural Road, Vancouver, BC V6T 1Z1 Canada}
\item{Lane Department of Computer Science and Electrical Engineering, 1220 Evansdale Drive, PO Box 6109, Morgantown, WV 26506, USA}
\item{Center for Gravitational Waves and Cosmology, West Virginia University, Chestnut Ridge Research Building, Morgantown, WV 26505, USA}
\item{Department of Electrical Engineering, Universidad de Chile, Av. Tupper 2007, Santiago 8370451, Chile}
\item{Cornell Center for Astrophysics and Planetary Science, Cornell University, Ithaca, NY 14853, USA}
\item{Department of Statistical Sciences, University of Toronto, 700 University Avenue 9th Floor, Toronto, ON, M5G 1X6, Canada}
\item{Department of Physics and Astronomy, West Virginia University, P.O. Box 6315, Morgantown, WV 26506, USA}
\item{CSIRO Space \& Astronomy, Parkes Observatory, P.O. Box 276, Parkes NSW 2870, Australia}
\item{Department of Astronomy, University of California Berkeley, Berkeley, CA 94720, USA}
\item{NASA Hubble Fellowship Program (NHFP) Einstein Fellow}
\item{Department of Astrophysical Sciences, Peyton Hall, Princeton University, Princeton, NJ 08544, USA}
\item{Department of Astronomy and Astrophysics, University of California Santa Cruz, Santa Cruz, CA 95064, USA}
\item{Kavli Institute for the Physics and Mathematics of the Universe (Kavli IPMU), 5-1-5 Kashiwanoha, Kashiwa,
277-8583, Japan}
\item{Division of Science, National Astronomical Observatory of Japan, 2-21-1 Osawa, Mitaka, Tokyo 181-8588, Japan}
\item{Laboratoire de Physique et Chimie de l’Environnement et de l’Espace - Universit\'e d’Orl\'eans/CNRS, 45071, Orl\'eans Cedex 02, France}
\item{Department of Physics and Astronomy, York University, 4700 Keele Street, Toronto, Ontario, ON MJ3 1P3, Canada}
\item{Perimeter Institute for Theoretical Physics, 31 Caroline Street N, Waterloo, ON N25 2YL, Canada}
\item{Ioffe Institute, 26 Politekhnicheskaya St., St. Petersburg 194021, Russia}
\item{Data Sciences Institute, University of Toronto, Ontario Power Generation Building, 700 University Ave., 17th Floor, M5G 1Z5}
\item{Anton Pannekoek Institute for Astronomy, University of Amsterdam, Science Park 904, 1098 XH Amsterdam, The Netherlands}
\item{ASTRON, Netherlands Institute for Radio Astronomy, Oude Hoogeveensedijk 4, 7991 PD Dwingeloo, The Netherlands}
\end{affiliations}
}
\newcommand{\allacks}{
A.B.P. is a Banting Fellow, a McGill Space Institute~(MSI) Fellow, and a Fonds de Recherche du Quebec -- Nature et Technologies~(FRQNT) postdoctoral fellow.
A.M.C. is funded by an NSERC Doctoral Postgraduate Scholarship. 
A.P. is funded by the NSERC Canada Graduate Scholarships -- Doctoral program.
A.P.C is a Vanier Canada Graduate Scholar.
B.C.A. is supported by an FRQNT Doctoral Research Award.
C.D.K. is partly supported by a CIERA postdoctoral fellowship.
C.L. is supported by NASA through the NASA Hubble Fellowship grant HST-HF2-51536.001-A awarded by the Space Telescope Science Institute, which is operated by the Association of Universities for Research in Astronomy, Inc., under NASA contract NAS5-26555.
D.C.S. is supported by an NSERC Discovery Grant (RGPIN-2021-03985) and by a Canadian Statistical Sciences Institute (CANSSI) Collaborative Research Team Grant.
D. Z. L is a  Lyman Spitzer, Jr.  Fellow
F.A.D is funded by the UBC Four Year Fellowship
G.M.E. acknowledges the financial support from an NSERC Discovery Grant (RGPIN-2020-04554) and from a Collaborative Research Team Grant from the Canadian Statistical Sciences Institute (CANSSI, with support from NSERC).
FRB research at UBC is supported by an NSERC Discovery Grant and by the Canadian Institute for Advanced Research. The baseband recording system for CHIME/FRB is funded in part by a CFI John R. Evans Leaders Fund award to I.H.S.
K.M.B. is supported by NSF grant 2018490
K.N. is an MIT Kavli Fellow. 
K.R.S is supported by an FRQNT Doctoral Research Award.
K.S. is supported by the NSF Graduate Research Fellowship Program.
K.W.M. holds the Adam J. Burgasser Chair in Astrophysics and is supported by NSF grants (2008031, 2018490).
M.B is a McWilliams fellow and an International Astronomical Union Gruber fellow. M.B. also receives support from the McWilliams seed grant.
The Dunlap Institute is funded through an endowment established by the David Dunlap family and the University of Toronto. B.M.G. acknowledges the support of the Natural Sciences and Engineering Research Council of Canada (NSERC) through grant RGPIN-2022-03163, and of the Canada Research Chairs program.
T.E. is supported by NASA through the NASA Hubble Fellowship grant HST-HF2-51504.001-A awarded by the Space Telescope Science Institute, which is operated by the Association of Universities for Research in Astronomy, Inc., for NASA, under contract NAS5-26555. 
V.M.K. holds the Lorne Trottier Chair in Astrophysics \& Cosmology and a Distinguished James McGill Professorship and receives support from an NSERC Discovery Grant and Herzberg Award, from an R. Howard Webster Foundation Fellowship from the Canadian Institute for Advanced Research (CIFAR), and from the FRQNT Centre de Recherche en Astrophysique du Quebec.
Z.P. was a Dunlap Fellow and is supported by an NWO Veni fellowship (VI.Veni.222.295).
}

%% file: methods.tex
\begin{methods}
\renewcommand{\figurename}{Extended Data Figure}
\setcounter{figure}{0}
\renewcommand{\tablename}{Extended Data Table}
\setcounter{table}{0}

\section*{Observations and burst detection}
On 2022-10-22 at $\sim$10:14 UTC, CHIME/FRB detected FRB 20221022A. The event was sufficiently bright to be detected across five of CHIME/FRB's real-time FFT search beams\cite{chime/frb2018}, attaining an S/N of 64.9 in the brightest beam and triggering the recording of channelized voltage (baseband) data at time/frequency resolution of 2.56 $\mu s$/0.391 MHz from each of CHIME's 1024 dual-polarised feeds. FRB 20201022A was initially identified as an interesting source due to its low excess DM, indicating the source as extragalactic but originating in the nearby Universe. Subsequent processing with the CHIME/FRB's baseband pipeline\cite{Michilli2021} provided the beamformed data on which polarisation analysis, reported here, could be applied.

\section*{Burst properties}
\subsection{Fitburst}
\begin{figure}[h]%
\centering
\includegraphics[width=0.95\textwidth]{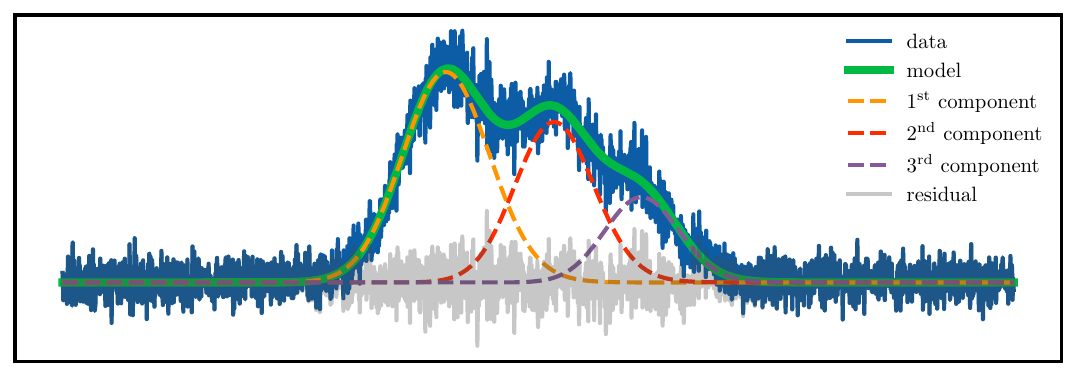} 
\caption{Frequency summed burst profile. We fit a three-component model to the dynamic spectrum of the data.}\label{fig:profile}
\end{figure}
We model the 2D dynamic spectrum of the burst using an extension of the \texttt{fitburst} fitting algorithm and use a GPU accelerated MCMC backend for parameter inference\cite{Giri:2023krc, Fonseca:2023nzl, Masui2015}. We fit a three-component burst model to the data (Stokes $I$) where the intrinsic burst profile is modelled as a Gaussian profile with a three-parameter power-law spectrum where the three parameters describe the amplitude, spectral index and spectral running. We assume a thin-screen model for scattering and model it using a decaying exponential with a scattering index of -4. We fix dispersion index to -2. The intrinsic parameters like time of arrival (ToA), width and spectrum are modelled separately for each of the three components whereas the propagation effects like dispersion measure and scattering time of the burst, which are expected to be same for each of the bursts, are modelled jointly. We assume a Gaussian likelihood model and use broad flat priors on our model parameters for defining the posterior. The posterior is sampled using the NUTS algorithm implemented in the numpyro library. The median based best fit point estimates for DM and scattering together with ToA and width of the sub-components are presented in Table\,\ref{tab1}. The frequency summed data and model corresponding to the best-fit point estimate is shown in Extended Data Fig.~\ref{fig:profile}.
\subsection{Scintillation measurements}
\label{meth:scintillation}
We measure two scintillation scales in the spectrum of FRB~20221022A: 9\,kHz and 194\,kHz at 600\,MHz (Nimmo et al. in preparation). Using the generalised form of Equation 6 in ref.\cite{occ+22}, we find that having both screens residing within the Milky Way would not have preserved the coherence in order to detect the scintillation. We therefore conclude that the second screen, and therefore the source, must be extragalactic, further supporting the host galaxy association. 
The NE2001 prediction is $\sim45$\,kHz at 600\,MHz (using the scattering timescale reported in Table\,\ref{tab1} and the relationship $\tau\sim1/(2\pi\nu_{\text{scint}})$; ref.\cite{cordes2002ne2001}). Both measured scales are roughly a factor of $5$ from the Galactic expectation, therefore it is difficult to conclude which of the two scales is the host scintillation scale. 
A detailed description of the scintillation analysis and constraints will be presented in a forthcoming paper (Nimmo et al. in preparation). 

\subsection{Flux/fluence measurements}
The flux density of baseband data is calibrated with a daily observation of a persistent source\cite{chime/frb2023b}. To account for the different system and sky temperatures at the time of the FRB detection, the off-pulse region in the FRB profile is subtracted from the on-pulse region, independently for the two polarisations.
A model of the telescope's primary beam is then used to rescale the flux density for sources detected away from the zenith.
Finally, the two polarisations are summed together. The peak flux density is calculated as the maximum flux density in the profile, while the fluence is obtained by integrating the profile in time with a simple rectangle method\cite{chime/frb2023b}.

\section*{Localization and position accuracy}
Baseband data are used to localize the burst to sub-arcminute precision. As detailed in ref.\cite{Michilli2021}, a grid of beams is formed in the sky around an initial guess position for the source. A S/N value is calculated for each beam, creating a 2D map of the signal strength in this sky area. The S/N map is fitted with a 2D Gaussian function that approximates the beam response. The FRB position and its uncertainties are obtained through this fit. Ref.\cite{Michilli2021} use a sample of known sources to estimate systematic effects on CHIME/FRB localizations. The result of their analysis is included in the FRB position uncertainties that we report here. The baseband localization situates the source in the main lobe of CHIME's primary beam. This firmly ruled out the possibility that the event was a misclassified Galactic source in a sidelobe, an important sanity check.

\section*{Host galaxy identification and 
$\rm{P(O|x)}$ estimate}
We downloaded Pan-STARRS 3$\pi$ (PS1) $r$-band and $g$-band images\cite{Waters16} of the approximate localization region of FRB\,20221022A and centered on $\rm{RA}=48.628$, $\rm{Dec}=86.872$ degrees (J2000).  Within the error circle of the FRB, there is a galaxy catalogued in the NASA Extragalactic Database (NED)\cite{NED} as MCG+14-02-011 and with a $B$-band magnitude of $\approx16$~mag, but it is blended with an extremely bright foreground star TYC\,4624-565-1 with a PS1 $r$-band magnitude of 11.88~mag and $g$-band magnitude of 12.51~mag\cite{Flewelling16}.  We therefore subtract the star from the PS1 image and obtain an approximate $r$-band and $g$-band magnitude of MCG+14-02-011 as follows.  We first construct an empirical point-spread function (PSF) for the PS1 image using {\tt photutils}\cite{photutils} and stars within 3\arcmin\ of the galaxy.  Then assuming that the flux and centroid for the bright star have the same values reported in the PS1 catalog, we subtract the scaled PSF from that position in the PS1 3$\pi$ images.  We mask out pixels where the residual from the subtraction either yields a negative flux or is $>$20$\sigma$ times the root-mean square pixel value above the background flux.  Finally, we apply an elliptical aperture with semi-major axis 0.34\arcmin\ and semi-minor axis 0.17\arcmin\ and a position angle of 165$^{\circ}$ east of north, which encapsulates $>$95\% of the galaxy flux in the PS1 images to estimate its brightness, with a background elliptical annulus with the same orientation and an inner radius 2$\times$ and outer radius 4$\times$ these values. To account for the masked pixels, we rescale our flux and uncertainty by the fraction of masked pixels within the elliptical aperture. Using these methods, we estimate that the host galaxy has a $r$-band brightness of 15.1$\pm$0.3~AB~mag and a $g$-band brightness of 15.7$\pm$0.2~AB~mag, which is approximately consistent with its $B$-band magnitude as reported by NED.

\begin{figure}[h]%
\centering
\includegraphics[width=0.6\textwidth]{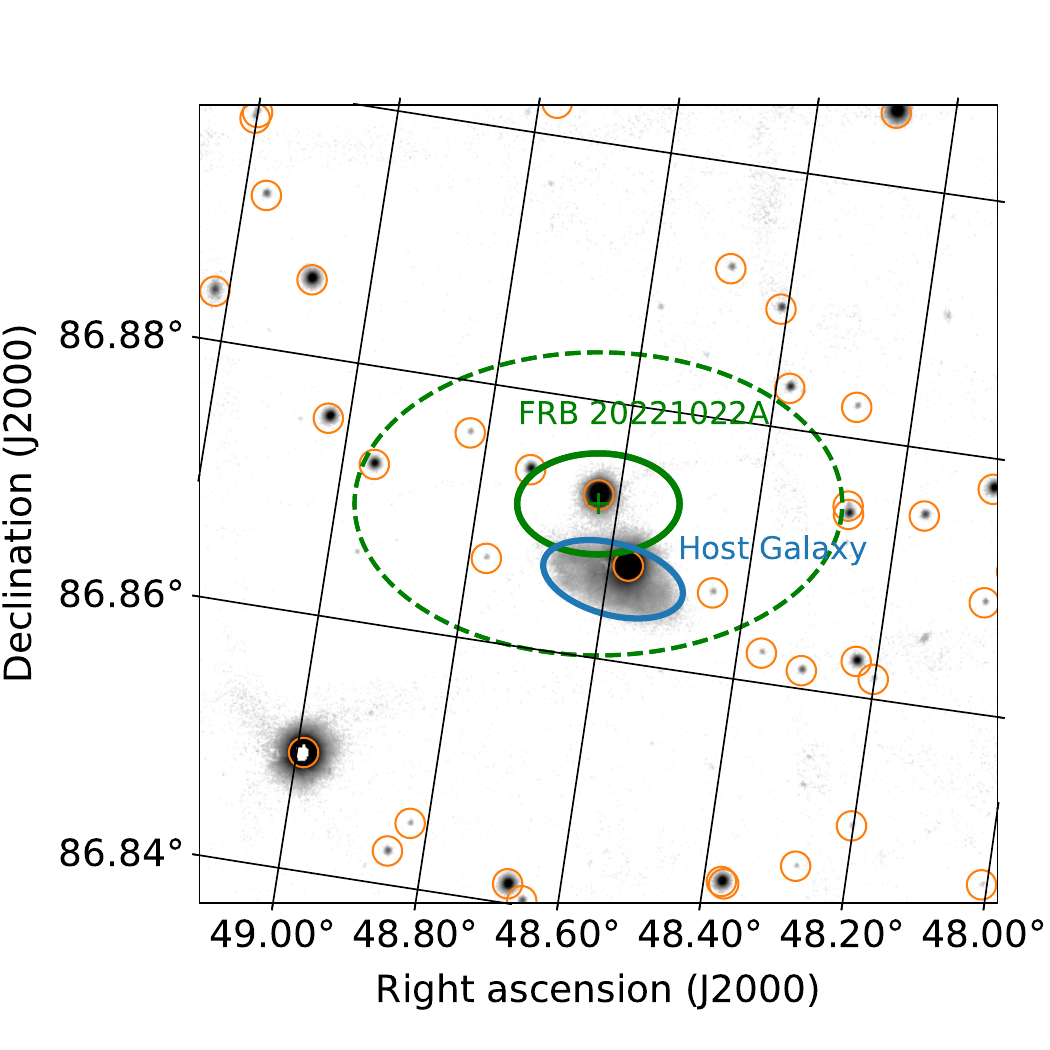} 
\caption{The same image as that displayed in panel (e) of Extended Data Fig.~\ref{fig:waterfall} but without any masking of the bright foreground star, TYC\,4624-565-1.}\label{fig:cutout_unmasked}
\end{figure}

To determine the association probability between FRB\,20221022A and the putative host, we utilize PATH (Probabilistic Association of
Transients to their Hosts)\cite{Aggarwal2021}, a Bayesian framework for transient host associations. We specify the prior probability that the host is unseen/undetected to be 0.2, (i.e., P(U)=0.2, a 20\% prior probability that the host is undetected) to account for the limiting magnitude of the PS1 image and an exponential offset prior with a scale length of 0.5. From our PATH analysis, we determine a posterior probability of $P(O|x) \gtrsim 99\%$ for MCG+14-02-011, and thus conclude the association is robust.  

We obtained the spectrum of the galaxy using the Optical System for Imaging and low-intermediate Resolution Integrated Spectroscopy (OSIRIS) instrument installed at the Gran Telescopio Canarias (GTC; program GTCMULTIPLE2C-22BMEX). The observations were performed on January 27, 2023, under clear conditions with seeing of about 1.5 arcsec. We acquired 72$\times$33s exposures using the R500B grism which covers the spectral range from 3600 \AA  ~to 7200 \AA. The spectra were obtained with the 1.5 arcsec slit width, providing a spectral resolution of about 20 \AA. The slit was oriented as shown in Extended Data Fig.~\ref{fig:spectroscopy}, with a $\rm{P.A.}=20^{\circ}$, to avoid contamination from the foreground star TYC\,4624-565-1. 

We performed standard data reduction using IRAF routines. The wavelength solution was achieved using a set of HgAr and Ne arc lamps, with the resulting rms $<$2\AA. The flux was calibrated using the G191-B2B spectrophotometric standard\cite{Oke1974, Massey1988, Oke1990} observed during the same night as the target galaxy. The resulting calibrated spectrum is shown in Extended Data Fig.~\ref{fig:spectroscopy}. Based on the the H$\alpha$, H$\beta$, [OIII], [NII], and [SII] line features, we estimate the spectroscopic redshift $z_{\rm spec}$ of the host to be 0.0149(3).

\begin{figure}
\begin{center}
\includegraphics[width=0.8\textwidth]
{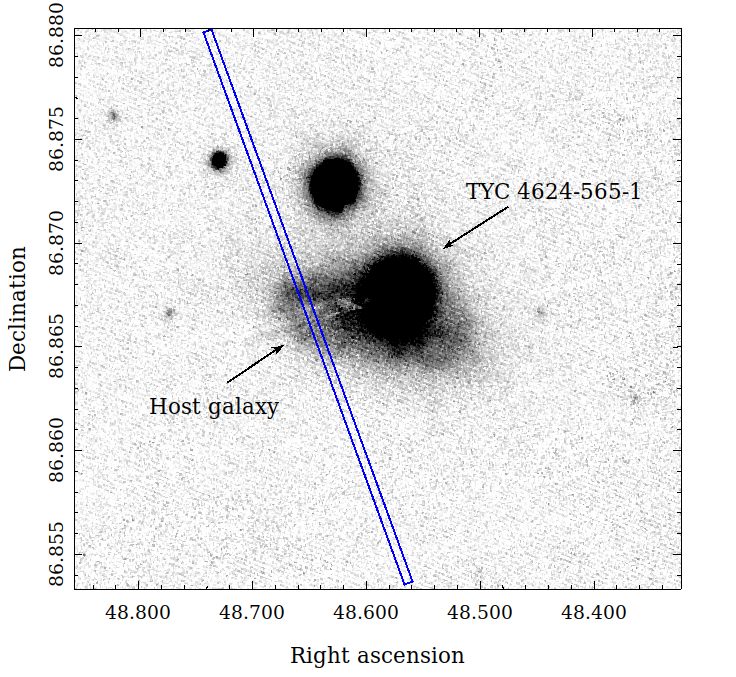} \\
\includegraphics[width=0.99\textwidth]
{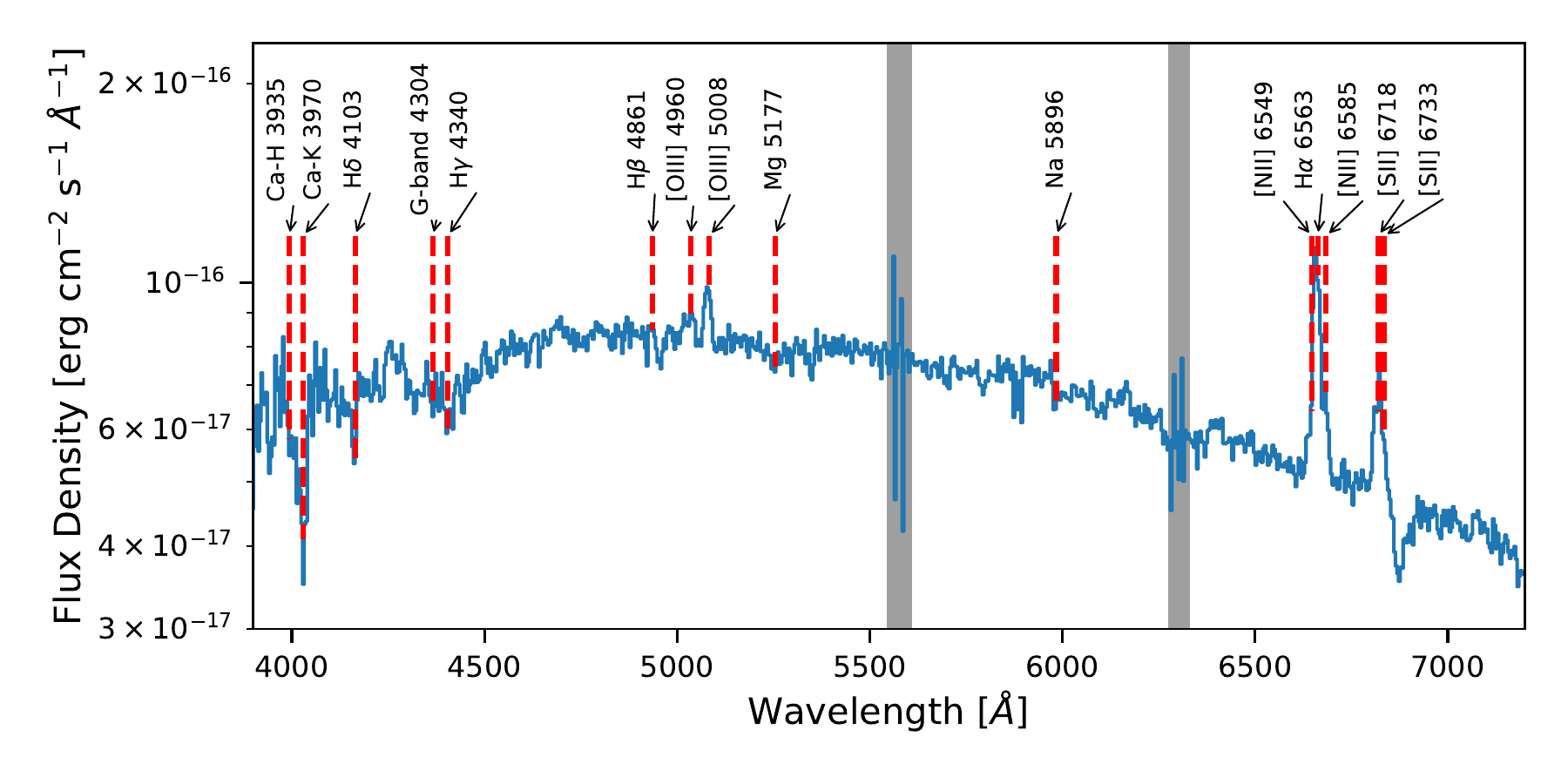} \\
\caption{Top: Pan-STARRS 3$\pi$ (PS1) $g$-band image of the FRB 20221022A field. Arrows indicate the likely FRB host galaxy and the foreground star. Blue rectangle shows the position and orientation of the slit in the GTC observations. Bottom: Optical spectrum of MCG+14-02-011 covered by the slit in the GTC observations, identified as likely host of FRB 20221022A.The two grey shaded regions mask artefacts in the spectrum.}\label{fig:spectroscopy}
\end{center}
\end{figure}  

To estimate the stellar mass (M$_{\ast}$) of MCG+14-02-011 we follow ref.\cite{bernardi2010MNRAS}. The mass$-$luminosity relation at redshift $z \approx$ 0 is given as a function of (g-r)$_{0}$ colors,
\begin{equation}
\rm log_{10}(M_{\ast}/L_{r}) = 1.097(g-r)_{0} - 0.406,
\end{equation}
where L$_{\rm r}$ = 10$^{7.8}$ L$_{\odot}$ is the r-band luminosity of the host calculated using $z$ = 0.0149 and m$_{\rm r}$ = 15.30 AB mag, and the factor $-0.406$ is derived by ref.\cite{bernardi2010MNRAS} using the Chabrier initial mass function \cite{chabrier2003PASP}. Using the Galactic extinction corrected (g-r)$_{0}$ color = 0.35, we estimate M$_{\ast}$ $\sim$ 10$^{9.6}$ M$_{\odot}$.   

\section*{Search for persistent radio sources \& high-energy counterparts}

We searched archival radio data from several surveys and identified a persistent radio source within the FRB localization region. The surveys analysed include the Tata Institute of Fundamental Research Giant Metrewave Radio Telescope Sky Survey Alternative Data Release\cite{intema2017gmrt}, the Westerbork Northern Sky Survey\cite{rengelink1997westerbork}, the NRAO VLA Sky Survey (NVSS)\cite{condon1998nrao}, and the VLA Sky Survey (VLASS)\cite{lacy2016vla}. Our search revealed only one radio source, NVSS J031417+865200, which is exclusively detected in NVSS and appears either unresolved or marginally extended. Moreover, it is spatially coincident with the center of the FRB host galaxy.

The integrated flux density of the NVSS source was estimated using the Aegean package\cite{2012MNRAS.422.1812H,2018PASA...35...11H} to be 1.5 $\pm$ 0.3 mJy. Based on the non-detection of the radio source in the VLASS survey, we estimate 3$\sigma$ upper limit on the flux density at 3 GHz to be 0.390 mJy. Using this upper limit and assuming a power-law dependence of the NVSS radio source flux density (i.e., $S_{\nu}$\,$\propto$\,$S^{\alpha}$), we estimate a lower limit on $\alpha$\,$<$\,--1.7, supporting its extended nature. NVSS J031417+865200 is likely the result of ongoing star formation in the FRB host galaxy. We estimate the SFR using the NVSS 1.4 GHz continuum emission. Using the 1.4 GHz-SFR relation from\cite{2011ApJ...737...67M}, SFR$_{\rm 1.4 GHz}/\rm{M_{\odot} yr^{-1}}$ = $\rm 6.35 \times 10^{-29} \times L_{1.4 GHz}/erg s^{-1} Hz^{-1}$, we calculate SFR$_{\rm 1.4 GHz} \sim 0.47 \rm{M_{\odot} yr^{-1}}$. The calculated SFR rate aligns well with the value derived using the main sequence SFR$-$stellar mass relation\cite{2004MNRAS.351.1151B}, providing evidence for the FRB host being a star-forming galaxy, consistent with the observed emission lines in the optical spectrum of the host.

We also carried a search for high-energy counterparts within the 3$\sigma$ localization region of FRB~20221022A. We searched the X-ray master catalog for X-ray sources with positions that are consistent with the location of FRB~20221022A. There are three X-ray sources \break (2RXS~J032618.0+865610, 2RXS~J025917.9+865216, and 1RXS~J033107.3+865052) located within 10-15 arcmin of FRB~20221022A's position. Since the positions of all of these X-ray sources are outside of FRB~20221022A's 3$\sigma$ localization region, we do not associate them with the FRB source. We also did not find any gamma-ray sources at the location of FRB~20221022A in any of the Fermi-LAT catalogues. The nearest catalogued gamma-ray point source is \break 4FGL~J0140.6+8736, which is significantly offset by 1.3$^\circ$ from FRB~20221022A. 

\section*{Polarisation analysis \& PA curve fitting}
The polarisation analysis used here follows a similar procedure to that previously applied to other CHIME-detected FRBs\cite{Mckinven2023,Mckinven2023b}
The complex, beamformed baseband data are converted to Stokes $I,Q,U,V$ parameters through the usual transformations\cite{Mckinven2021}. Stokes $I,Q,U,V$ spectra are extracted by integrating the signal over only the leading burst component of FRB 20221022A (indicated by vertical gray dotted lines in the waterfall plots of Extended Data Fig.~\ref{fig:waterfall_QU}).
This was done to limit the deleterious effects of the substantial PA wrapping across the burst envelope. The resulting Stokes spectrum was then fit ($QU$-fitting) with a nested sampling routine that fits an in-house model that accounts for both astrophysical parameters such as the linear/circular polarisation fractions and RM as well as first order instrumental polarisation effects known to be present in CHIME observations\cite{Mckinven2021}. With this model we determine an RM and use this to unwrap the Stokes $Q,U$ signal. The lower panel of Extended Data Fig.~\ref{fig:waterfall_QU} displays the resulting \textit{derotated} Stokes $Q,U$ parameters that have been corrected for the $\lambda^2$ modulation of Faraday rotation. These derotated Stokes parameters are then used to construct the PA as a function of time by integrating their signals over the full $400-800$ MHz band, resulting in $Q_{\rm{derot}}$, $U_{\rm{derot}}$ profiles from which the PA curve can be determined as,
\begin{equation}
\rm{PA} = \frac{1}{2}\tan^{-1}\frac{\textit{U}_{\rm{derot}}(t)}{\textit{Q}_{\rm{derot}}(t)}.
\label{eqn:PA}
\end{equation}
Here, PA measurements are not calibrated for absolute polarisation angle due to unknown effects from CHIME's primary beam and are shown relative to 0 degrees (panel (d) of Fig.~\ref{fig:waterfall}). 

\begin{figure}
\centering
\includegraphics[width=0.8\textwidth]{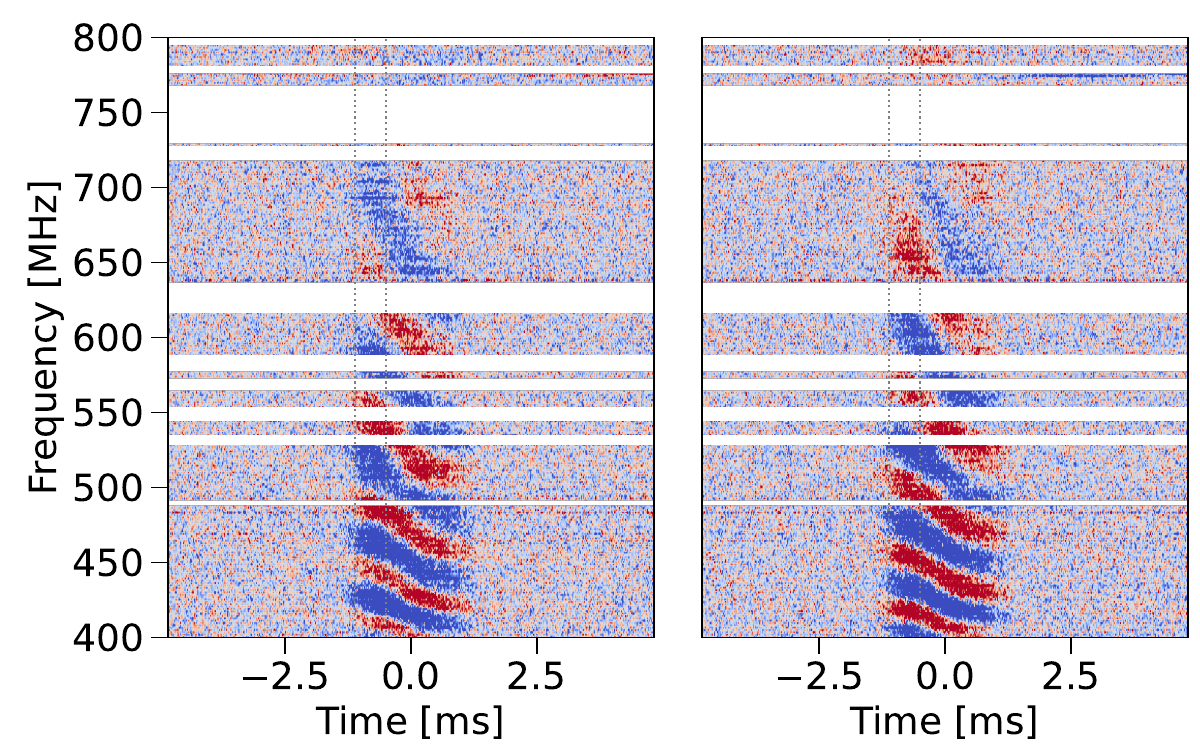} \\
\includegraphics[width=0.8\textwidth]{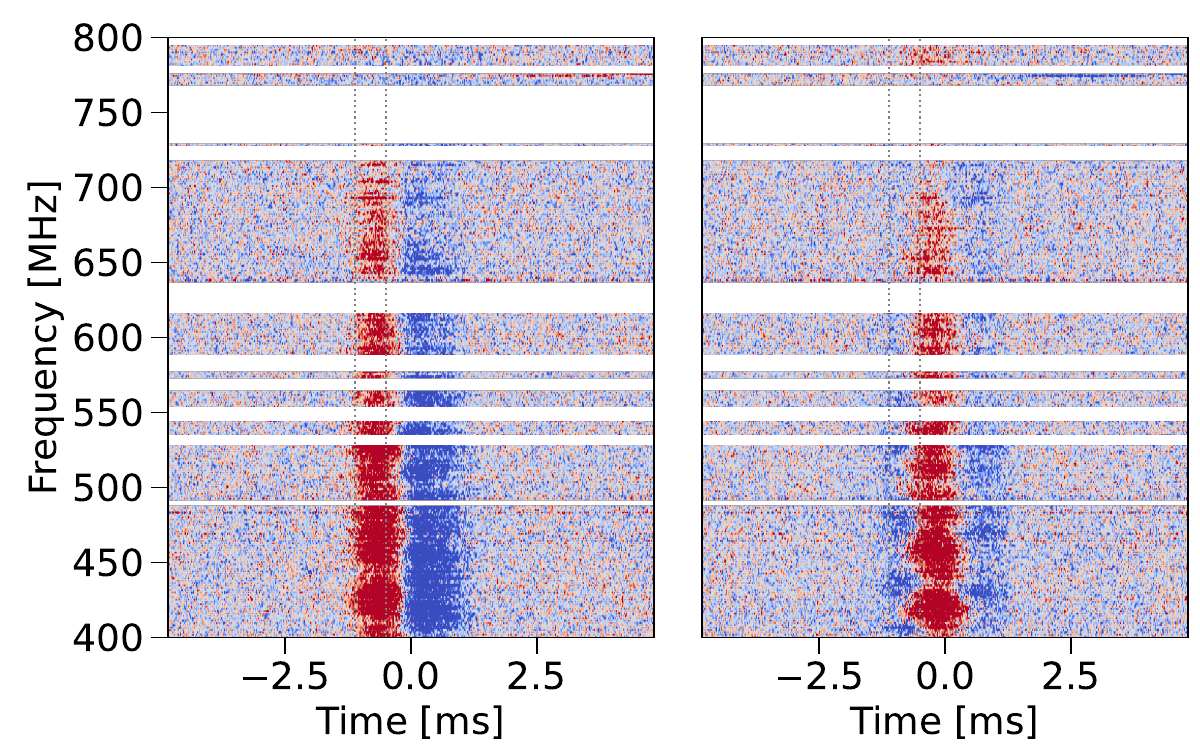}
\caption{Stokes $Q$ (left) and $U$ (right) waterfalls plots of FRB 20221022A before (top) and after (bottom) correcting for Faraday rotation. Both Faraday rotation and the PA swing are clearly evident in the ``candy cane" pattern imprinted on the Stokes $Q,U$ prior to correction (top panel). The faint $\sim30$ MHz ripple in the corrected Stokes $U$ waterfall is likely instrumental, an artefact of small differences in beam phase of CHIME's two linear polarisations. Displayed data have been rebinned in both time and frequency to resolutions $20.48\;\rm{\mu s}$ and 1.5625 MHz, respectively. White horizontal regions correspond to missing or masked frequency channels. Vertical grey dotted lines indicate the burst time limits that were used to construct the Stokes $I,Q,U,V$ spectrum.}\label{fig:waterfall_QU}
\end{figure}

\begin{figure}
\centering
\includegraphics[width=0.8\textwidth]{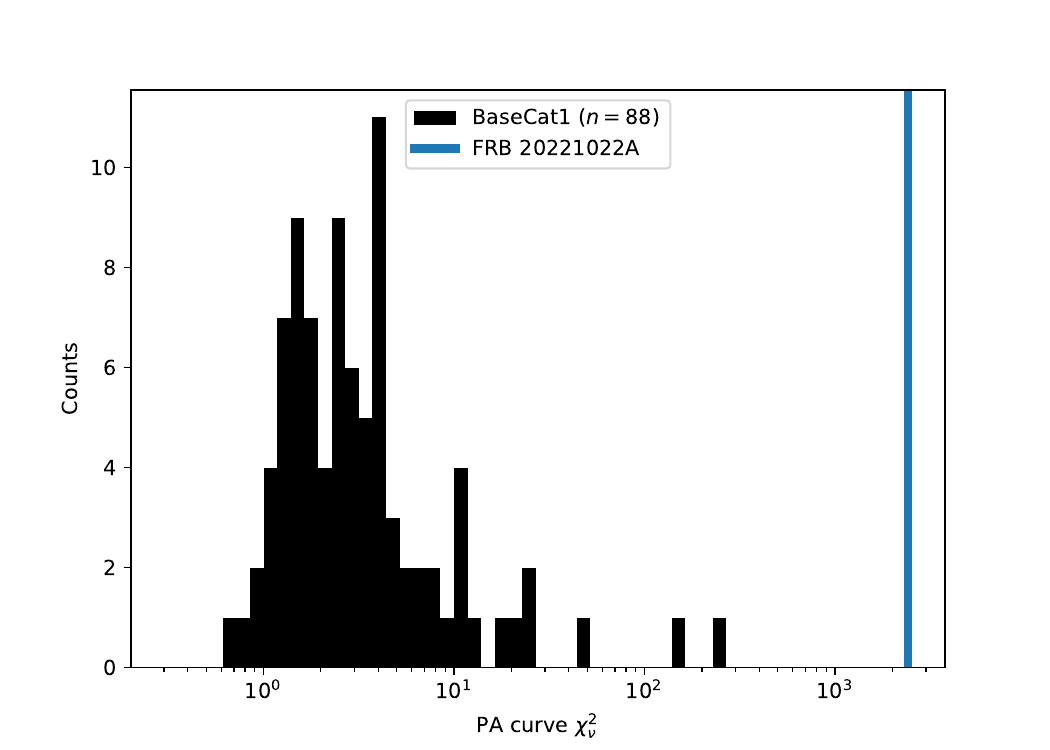}
\caption{A histogram of the reduced chi-square statistic measuring the PA curve variability, $\chi^2_{\nu}$, of the CHIME/FRB Baseband Catalog 1 (BaseCat1) polarized sample\cite{Pandhi2024}. The corresponding $\chi^2_{\nu}$ measurement of FRB 20221022A, represented by a vertical blue line, is noticeably larger than typical values of bursts detected by CHIME/FRB.) }
\label{fig:PA_curve_stats}
\end{figure}

The resulting PA curve shown in Fig.~\ref{fig:waterfall} is displayed at a time resolution of $20.48\mu s$ and only represents PA measurements where linearly polarised emission is above $5\sigma$. We measure the significance of the PA swing with a reduced chi-square statistic, $\chi^{2}_{\nu}$, which evaluates the deviation of the observed PA curve with respect to a constant (i.e., flat) PA = 0$^{\circ}$ model. FRB 20221022A's $\chi^{2}_{\nu}$ of 2413 is substantially larger than the distribution of $\chi^{2}_{\nu}$ values determined from a polarized subsample of one-off bursts ($n=88$ bursts) of the CHIME/FRB Baseband Catalog 1 sample\cite{Pandhi2024}. This comparison of the $\chi^{2}_{\nu}$ statistic is represented in Extended Data Fig.~\ref{fig:PA_curve_stats}, where the $\chi^{2}_{\nu}$ measurement of FRB 20221022A significantly is displaced relative to the BaseCat1 distribution; a factor of ten larger than the most extreme $\chi^{2}_{\nu}$ value of BaseCat1. 

We observe a low level residual circular polarisation even after correcting for dominant sources of instrumental leakage. The residual signal does not exceed $10\%$ across the majority of frequency channels and modulates in a manner similar to Faraday rotation which is consistent with expectations given known second order perturbations to the linear phase wrapping introduced from the fitted delay of the CHIME's X, Y polarised feeds\cite{Mckinven2021}. As such, the $\Pi_L$ measurements reported in Table~\ref{tab1} should be taken as lower limits on the linear polarisation and the true value is likely higher, approaching $\sim 100\%$. Polarimetric studies of pulsars have demonstrated a trend of deteriorating fit quality of the RVM in pulsars that display high fraction of circular polarisation compared to linear polarisation\cite{Johnston2023, Sobey2023}. Thus, the high linear fractional polarisation of FRB 20221022A and the success in the RVM in replicating the observed PA evolution squares nicely with these previous pulsar studies.


\begin{figure}
\centering
\includegraphics[width=0.99\textwidth]{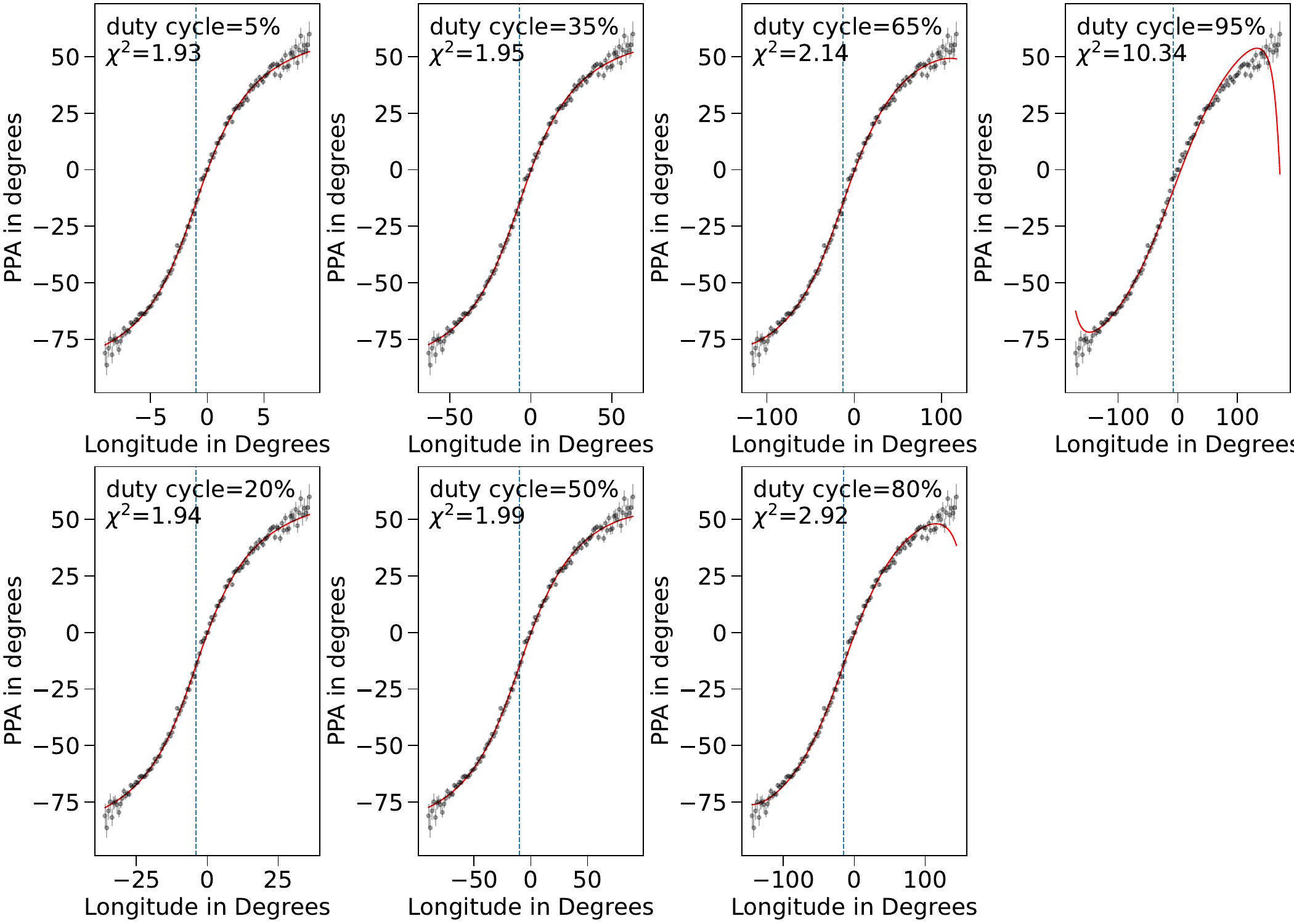}
\caption{PA curves (grey points) and associated best-fit RVM models (red) for a series of different assumed duty cycles (5, 20, 35, 50, 65, 80, 95$\%$). Quality of fits, indicated by reduced chi-squared values ($\chi^2$; top left corner), decreases at larger duty cycles indicating a general preference for models with smaller duty cycles and thus longer periods. Vertical blue lines indicate the phase ($\phi$ in Equation~\ref{eqn:rvm}) where the PA swing is steepest.}\label{fig:PA_curve_fits}
\end{figure}

Following a similar procedure to RVM fitting of pulsars\cite{Rookyard2015}, we perform a grid ($1000\times500$ bins) search over $\alpha$, $\beta$ space at trial duty cycles of 5\%, 20\%, 35\%, 50\%, 65\%, 80\%, 95\%. At each ($\alpha,\beta$) grid point, the least-$\chi^2$ fit between the RVM curve and the data is determined by optimising the remaining free parameters $\phi_{0}$ and $\psi_{0}$. The $\chi^2$ values resulting from the grid search were used to determine the likelihood regions displayed in Fig.~\ref{fig:alpha_beta}, representing $1\sigma$ and $3\sigma$ uncertainties. These uncertainties were calculated by following the standard procedure of determining confidence intervals from $\chi^2$ distributions across multiple dimensions\cite{Avni1976},
namely, at each period/duty cycle trial the minimum $\chi^2$ value, $\chi_{\rm{min}}^2$, was determined and 1$\sigma$ \& 3$\sigma$ contours were calculated as $\chi_{\rm{min}}^2+2.3$ and $\chi_{\rm{min}}^2+9.21$, respectively.  

Extended Data Fig.~\ref{fig:PA_curve_fits} shows the best-fit RVM PA curves at each trial duty cycle. An evident deterioration in fit quality is seen at larger duty cycles ($\gtrsim 50\%$), where the RVM struggles to reproduce the shallow PA gradient at the burst edges. At low duty cycles $\lesssim 50\%$ the RVM fit quality appears to converge toward a reduced chi-square value below 2. A second grid search was conducted for a sample of small duty cycles ($2.5, 1.0, 0.5, 0.1\%$) by constraining the grid to $\beta\leq 2.5$ degrees in order to better resolve the chi-square dependence. The resulting best-fit RVMs are not represented in Extended Data Fig.~\ref{fig:PA_curve_fits} but are consistent with fit quality obtained at larger duty cycles.

\section*{Constraints on beaming angle and implications for burst energy}

\begin{figure}
\centering
\includegraphics[width=1.1\textwidth]{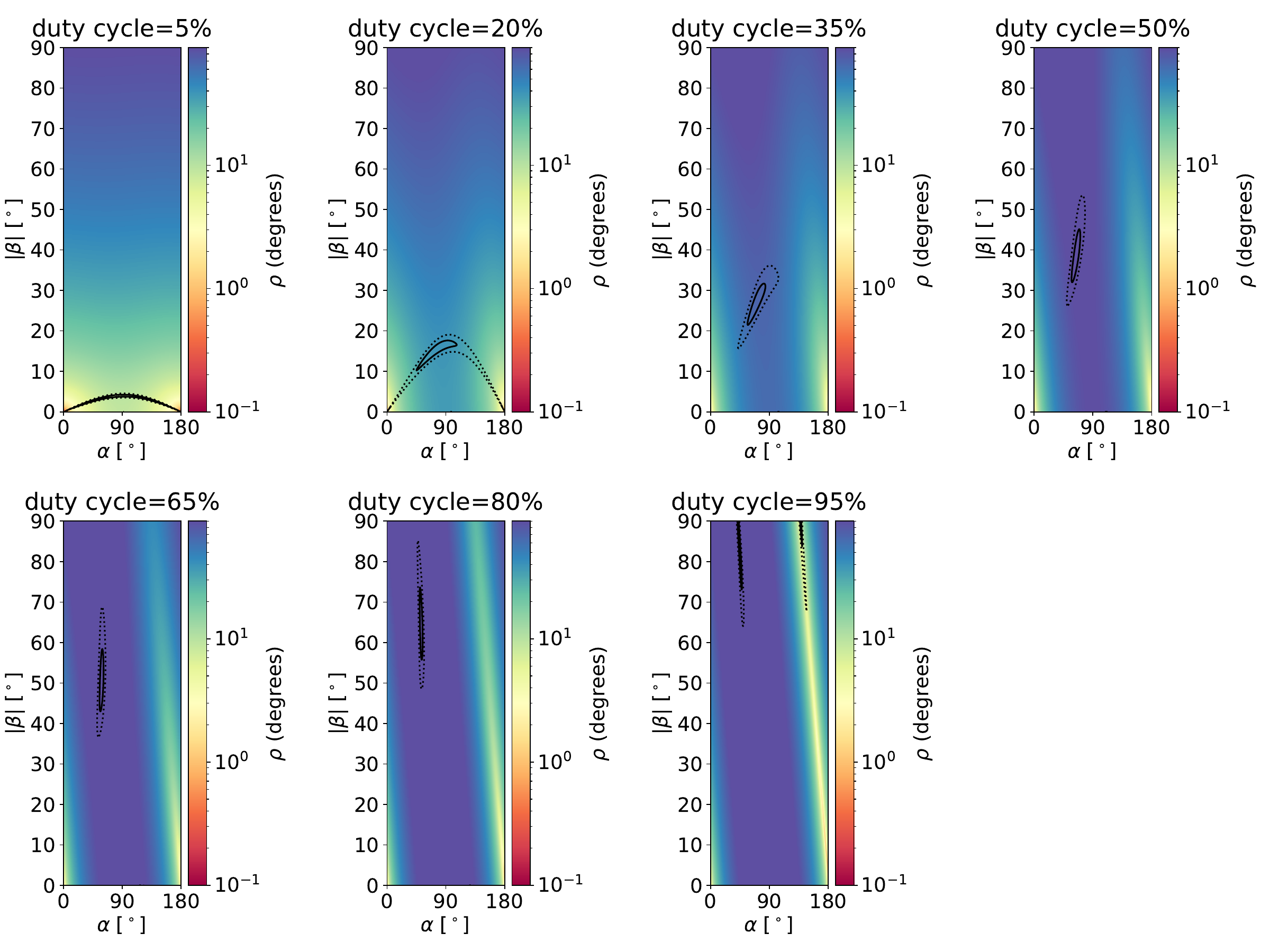}
\caption{Constraints on the RVM $\alpha, \beta$ parameter phase space for different assumed duty cycles of FRB 20221022A. For each panel, the beaming angle, $\rho$, is determined from Equation~\ref{eqn:rho} and is represented by a logarithmic (base 10) color map. The $1\sigma$ (solid line) \& $3\sigma$ (dotted line) contour lines for best-fit $\alpha, \beta$ values indicate a general trend toward smaller inferred beaming angles for smaller assumed duty cycles.}
\label{fig:alpha_beta_vs_rho}
\end{figure}

\begin{figure}
\centering
\includegraphics[width=0.99\textwidth]{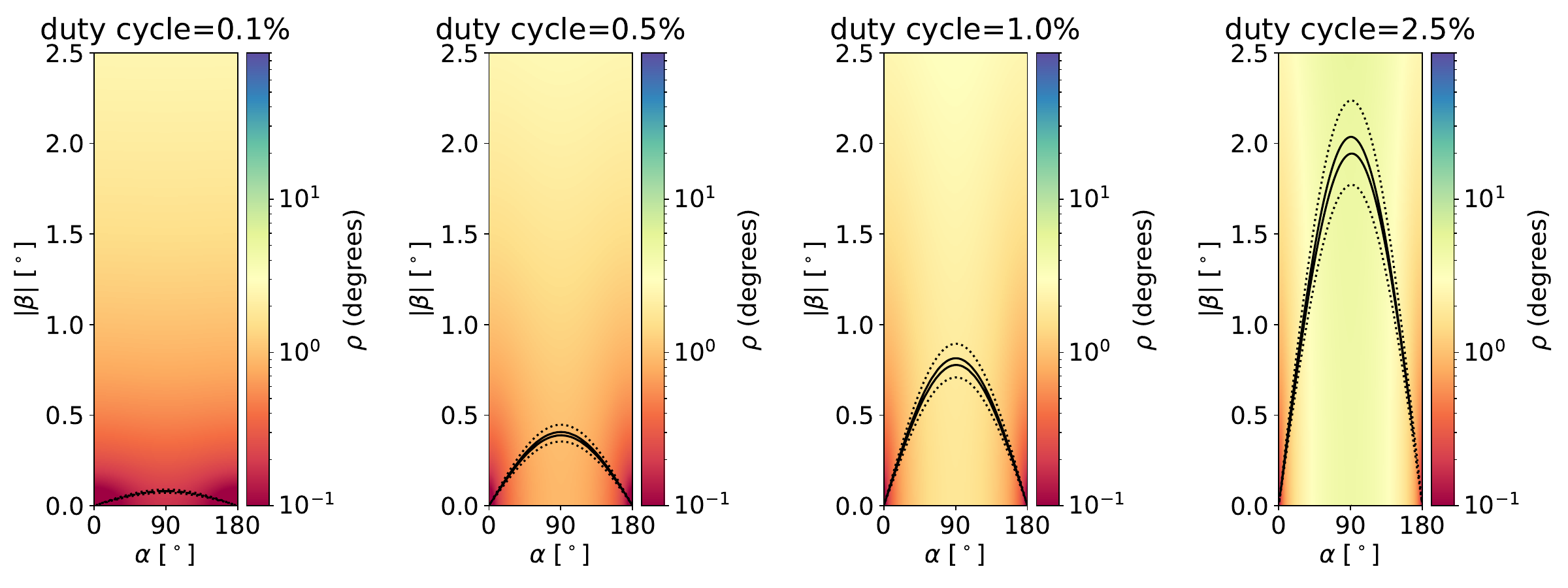}
\caption{Constraints on the RVM $\alpha, \beta$ parameter phase space for a sample of small trial duty cycles ($2.5,1.0,0.5,0.1\%$) that represents a continuation of a general trend toward smaller inferred beaming angles ($\rho$) for smaller assumed duty cycles represented in Extended Data Fig.~\ref{fig:alpha_beta_vs_rho}. Here, the $\beta$ axis has been confined to $\beta<2.5^{\circ}$ to resolve best-fit $\alpha,\beta$ contours which cluster preferentially at low $\beta$ values when duty cycles are small.}
\label{fig:alpha_beta_vs_rho2}
\end{figure}

If we assume that the observed emission of FRB 20221022A is beamed in a manner similar to that of pulsars, then the total energy release of the FRB source may be dramatically lower than what would be inferred if the source emits isotropically. For pulsars, the half-opening angle, $\rho$, of a conical emission beam centred on the magnetic axis can be related to the geometric parameters $\alpha$, $\beta$ and $W_{\rm{open}}$, the range in pulse phase where the open-field-line region transits the LoS\cite{Gronkowski1984},

\begin{equation}
\cos({\rho})=\cos(\alpha)\cos(\alpha+\beta)+\sin(\alpha)\sin(\alpha+\beta)\cos\bigg(\frac{W_{\rm{open}}}{2}\bigg).
\label{eqn:rho}
\end{equation}

We measure the burst width at $10\%$ of peak intensity, $W_{10\%}=2.5(0.2)\; \rm{ms}$, for easy comparison to the Galactic pulsar sample for which burst widths are often reported in this way. By assuming a period of the source, our $W_{10\%}$ measurement can be converted into degrees of longitude, $W_{\rm{open}}$. Using these $W_{\rm{open}}$ measurements, along with best-fit $\alpha, \beta$ values, we can determine $\rho$ at each trial duty cycle and find an approximately linear trend of smaller beaming angles for smaller assumed duty cycles. Uncertainties on the $W_{10\%}$ measurement are determined from the time resolution, $t_{\rm{samp}}$, the off pulse RMS, $\sigma_I$, and the equation\cite{Kijak1997,Wang2023},
\begin{equation}
\sigma_{W_{H}} = t_{\rm{samp}} \sqrt{1+\bigg(\frac{\sigma_I}{H}\bigg)^2}.
\end{equation}
Here, $H=0.1$ is used to determine the uncertainty on the width at $10\%$ of the burst's peak. 

Since the RVM fitting takes place with respect to pulse phase, our determination of RVM-parameters $\alpha$, $\beta$ and derived quantities like $\rho$ is insensitive to cosmological time dilation and whether we measure $W_{10\%}$ in the observer or source frame. That said, time dilation will affect our joint constraints on the duty cycle and period of the source. However, FRB 20221022A is very nearby ($z\sim 0.015$) and the $\frac{1}{1+z}$ scaling required to transform into the source frame is insignificant relative to the estimated measurement uncertainty of $W_{10\%}$. The effect of time dilation can thus be considered negligible on our joint constraints on duty cycle and period phase space of FRB 20221022A (see panel (a) of Extended Data Fig.~\ref{fig:pulsar_compare}).

Using the fluence measurement reported in Table~\ref{tab1}, we calculate the equivalent isotropic burst energy over the entire $400-800$ MHz band using the equation\cite{Zhang2018},

\begin{equation}
E_{\rm{iso}} = (10^{39} \rm{erg})\frac{4\pi}{1+z}\bigg(\frac{D_L}{10^{28} \rm{cm}}\bigg)^2\bigg(\frac{F_{\nu}}{\rm{Jy}\cdot \rm{ms}}\bigg)\bigg(\frac{\nu_c}{\rm{GHz}}\bigg) = 2.62(33)\times 10^{38}\; \rm{erg}.
\label{eqn:burst_E}
\end{equation}
Here, $\rm{D_{L}}=65.2(1.3)$ Mpc represents the luminosity distance to FRB 20221022A's host galaxy, $\rm{F}_{\nu}=87(11)\; \rm{Jy\, ms}$ is the specific fluence determined from the beamformed baseband data of the burst and the central frequency of the burst ($\nu_c$), taken here to be 0.6 GHz  
\begin{figure}
\centering
\includegraphics[width=0.65\textwidth]{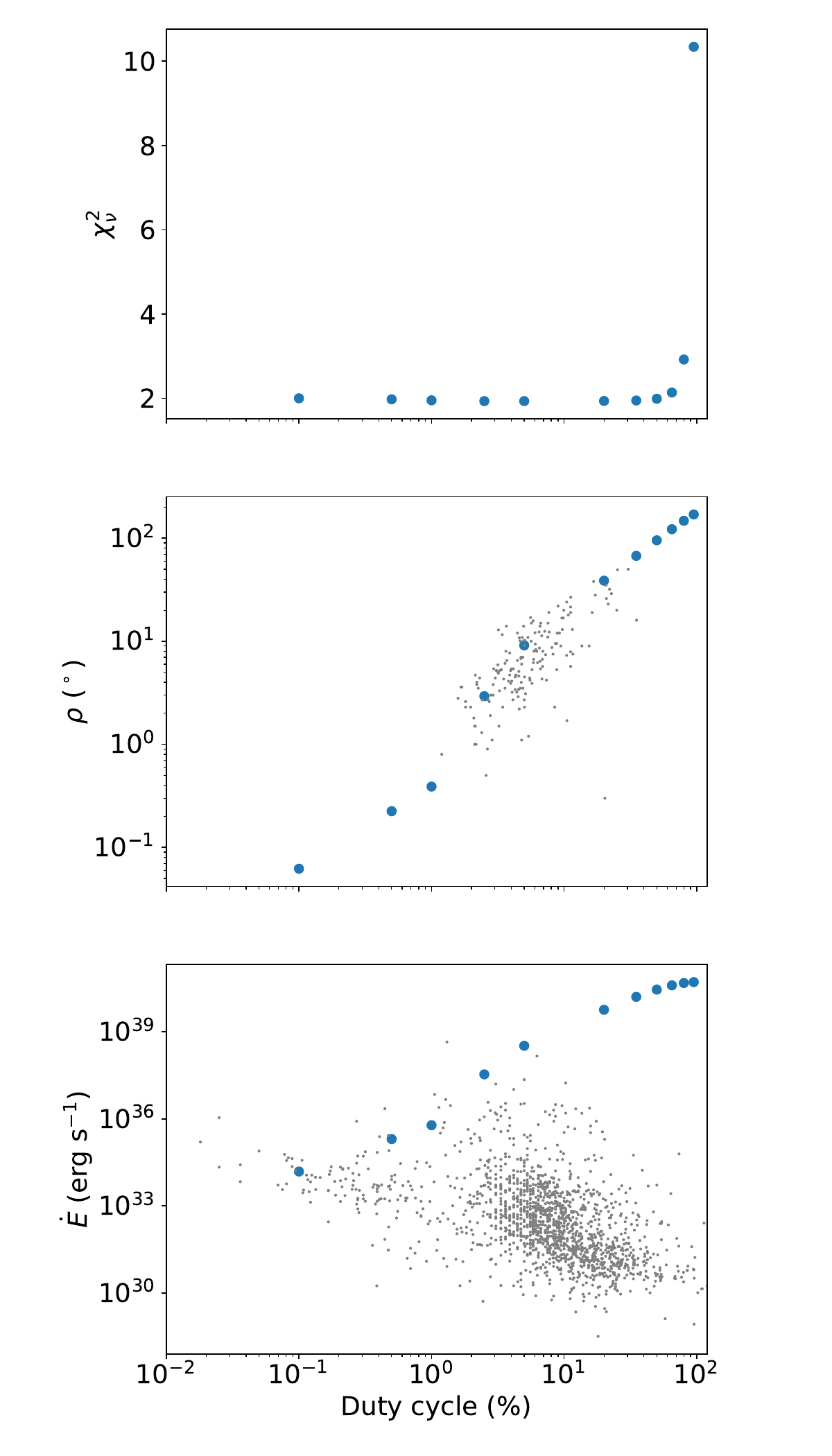}
\caption{Top panel: the reduced chi-squared ($\chi_{\nu}^2$) value of the best-fitting RVM model at trial duty cycles of 5, 20, 35, 50, 65, 80 \& 95$\%$. Middle panel: The inferred opening angle, $\rho$, calculated from Equation~\ref{eqn:rho} from best-fitting $\alpha,\beta$ RVM parameters and $\rm{W_{open}}$ measurement (Equation~\ref{eqn:rho}). Bottom panel: The implied spin-down burst luminosity, $\dot E$, corrected for the opening angle of the beam determined by rescaling the isotropic burst luminosity ($\rm{\dot E_{iso}}=1.05(16)\times10^{41}$ erg) by the the fraction of the solid angle subtended by $\rho$. Grey points represent equivalent spin-down luminosity measurements for a subsample of ATNF-catalog pulsars\cite{Wang2023}.}
\label{fig:rho_v_energy}
\end{figure}

If the emission from FRB 20221022A is highly beamed, then the total burst energy may be several orders of magnitude lower than would otherwise be required if the source emitted isotropically. The conical solid angle subtended by the $\rho$ is $\Omega = 2\pi(1-\cos(\rho))$, and the equivalent burst energy for this beamed emission can be calculated from by scaling $E_{\rm{iso}}$ by $\Omega/4\pi$. If we assume a beaming angle typical to pulsars, commonly $\lesssim 20$ degrees\cite{Wang2023}, then burst energies $\sim 3.7\times 10^{36}$ erg are sufficient to be detected at the observed fluence. Such energies are high relative to the general pulsar sample but not outside the range that could be provided by the spin-down luminosity of a young pulsar. This comparison, however, omits the radio emission efficiency factor, $\xi$, which for pulsars spans a wide range and is observed to be strongly correlated with characteristic age\cite{Szary2014} such that young pulsars typically display $\xi\lesssim 10^{-6}$. 

\section*{Comparison to Galactic pulsar sample}

A wide range of $\alpha,\beta$ combinations is plausible for FRB 20221022A given the uncertainty on the duty cycle/period of the source. However, since the burst width is known, an assumption on the duty cycle necessarily implies a period and vice versa. This inverse relation between assumed period and duty cycle is represented in panel (a) of Extended Data Fig.~\ref{fig:pulsar_compare} as a red line. This line is plotted alongside catalog values taken from the Australia Telescope National Facility (ATNF) pulsar catalogue\cite{Manchester2005}. Although the red line spans a large region of the phase occupied by the pulsar sample, at periods exceeding $\sim 200\; \rm{ms}$ it predicts duty cycles far below what is commonly observed from the pulsar sample, represented by a dashed grey line indicating an empirical boundary\cite{Johnston2019, Wang2023}, duty cycle $\propto P^{-1/3}$. If the source of FRB 20221022A is indeed a pulsar, then this region of phase space would appear to be disfavoured by the absence of any equivalent observations in known Galactic pulsars. Meanwhile, opening angle ($\rho$) measurements reported for a subsample of ATNF pulsars\cite{Wang2023} are displayed by a colour scale and indicate a trend of a narrowing beam for pulsars with smaller duty cycles. Pulsars at the low end of the duty cycle distribution regularly display $\rho$ values $\lesssim 5$ degrees, significantly smaller than the equivalent $\rho$ values obtained for FRB 20221022A at similar duty cycles (see Extended Data Fig.~\ref{fig:rho_v_energy}).     
We calculate the spin-down luminosity ($\dot E$; i.e., the presumed luminosity of magnetic dipole radiation inferred from the measured loss rate of rotational energy) from catalog $P,\dot P$ measurements and the equation,  
\begin{equation}
|\dot E| = \frac{4\pi^2I\dot P}{P^3}.
\label{fig:spindownE}
\end{equation}
Here, we assume a neutron star moment of inertia of $I=10^{45}\; \rm{g\,cm^2}$ and radiation efficiency of $100\%$ for the sake of illustration. Contours of constant spin-down luminosity represented in panel (b) of Extended Data Fig.~\ref{fig:pulsar_compare} and can be compared to the isotropic burst luminosity, $\rm{\dot E_{\rm{iso}}} = 1.05(16)\times 10^{41}\; \rm{erg\, s^{-1}}$, represented in $P\dot P$ space of panel (b) as a thick red line and determined from the intrinsic burst energy of the source and its temporal duration (see Methods). The orange diagonal line represents the magnetic field at the light cylinder, $B_{\rm{LC}}=10^5 \rm{G}$. Points lying above this line have larger $B_{\rm{LC}}$ values and are thus more likely to emit giant pulses\cite{Kuzmin2007}.

Comparison to the inferred spin-down luminosities of the Galactic pulsar sample indicates that the isotropic burst luminosity of FRB 20221022A is at least a few orders of magnitude larger than those of pulsars with the greatest inferred $\dot E$, a disparity that increases with the inclusion of a radio efficiency factor, $\xi$. This difference can be reduced if the emission from FRB 20221022A is highly beamed, however, the required opening angle would need to be $\lesssim 1$ degree. Such a value is not without precedent in the pulsar sample, however, in order to achieve such a small beaming angle, FRB 20221022A would need a duty cycle/period combination that is distinct from the pulsar sample, lying somewhere near or below the dashed line of panel (a) and a radio efficiency, $\xi$, much larger than what is seen from the Galactic pulsar sample. The absence of equivalent pulsar measurements in this region of duty cycle, period phase space seems to be at odds with FRB 20221022A having a small beaming angle. However, it may point towards a yet undetected population of long period, highly transient radio emitting neutron stars, similar to the suspected radio magnetar, PSR J1710-3452\cite{Surnis2023} and other ultra-long-period sources that have been recently discovered in our own Galaxy\cite{Caleb2022b,Hurley-Walker2023a, Hurley-Walker2023b}. Indeed, magnetar origins of FRBs are well-motivated by the highly luminous bursts observed from a Galactic magnetar\cite{Bochenek2020, chime/frb2020}, and the limited sample of radio-polarimetric observations have demonstrated magnetars capable of displaying many properties that are similar to what is reported here for FRB 20221022A, namely, very high $\Pi_L$ and S-shaped PA swings\cite{Lower2021}. Alternatively, the source may be a relatively short period but young pulsar with large spin-down luminosities, which like this FRB, tend to display large linear polarisation fractions. A color scale in panel (b) represents $\Pi_L$ measurements for a subsample of ATNF-catalog pulsars\cite{Wang2023}, which are significantly correlated with inferred spin-down luminosities. If FRB 20221022A does indeed originate from a young pulsar and its associated rotational energy loss, then the high level of polarisation ($\Pi_L\sim 100\%$) appears to agree with expectations for a source with an anomalously large $\dot E$.

\begin{figure}[h]
\centering
\includegraphics[width=0.55\textwidth]{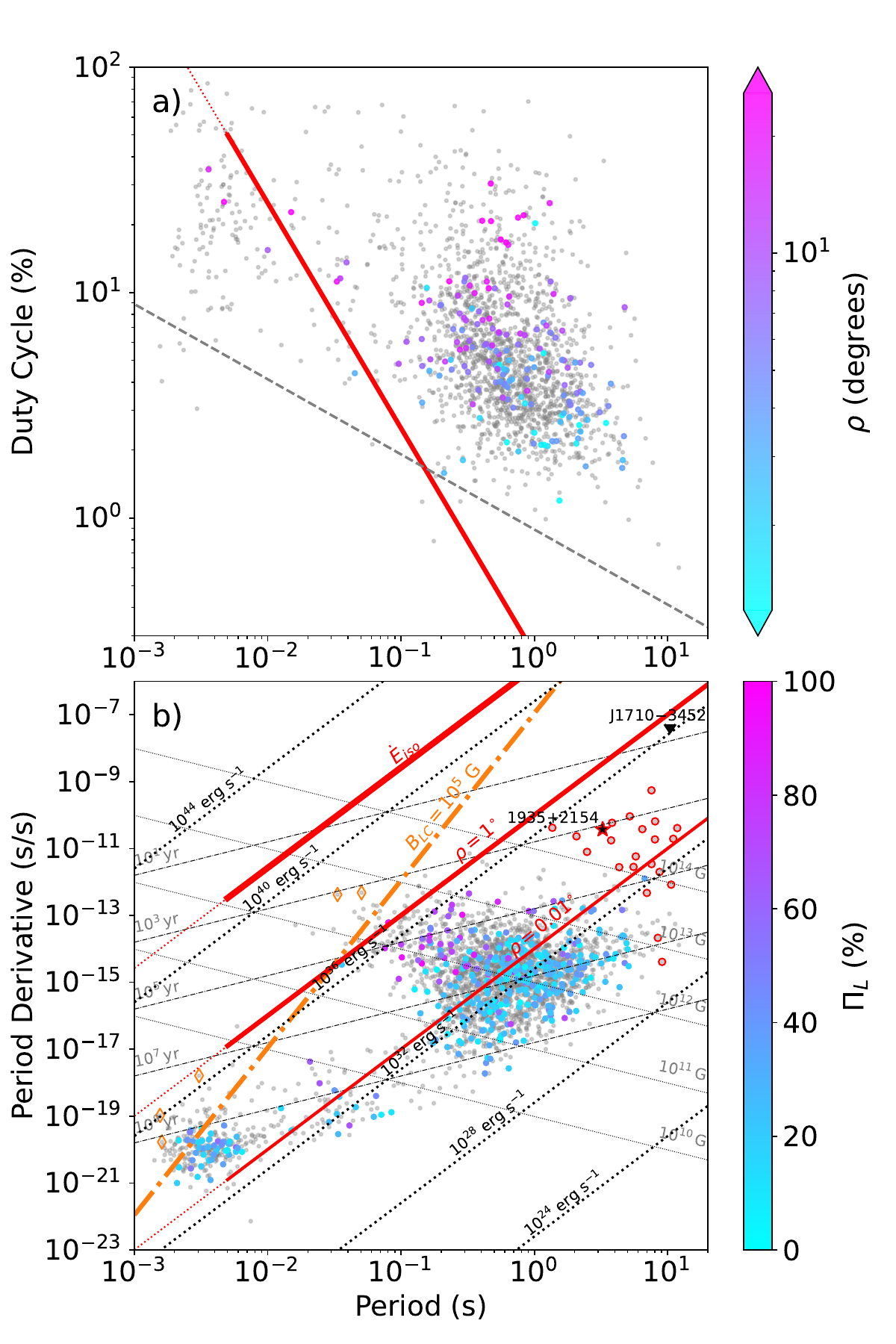}
\caption{Comparison of FRB 20221022A and selected parameters of the Galactic pulsar population. For both panels grey dots correspond to values taken from the Australia Telescope National Facility (ATNF) pulsar catalogue (version 1.70)\cite{Manchester2005}. a) Period versus duty cycle: a subsample with polarimetric constraints on the opening angle\cite{Wang2023}, $\rho$, are represented by a (logarithmic) color scale. The grey dotted line indicates an empirical lower boundary of the pulsar period-width relation\cite{Johnston2019}, $3.2\rm{P}^{-1/3}$. The red line indicates contraints on the combined period \& duty cycle of FRB 20221022A, with the dotted segment indicating the region of phase space excluded from RVM fitting ($\rm{period}\lesssim5$ ms; see text). b) Period and period derivative diagram ($\rm{P\dot P}$): dotted diagonal lines represent spin-down luminosities determined from Equation~\ref{fig:spindownE}. Red diagonal lines display the equivalent luminosity of FRB 20221022A for isotropic ($E_{\rm{iso}}$) or beamed emission ($\rho=(1,0.01)$ degrees). A subsample with measurements of the linear polarisation fraction, $\Pi_L$, are represented by a colour scale and indicate a tendency for highly linearly polarised events to occupy a region of $\rm{P\dot P}$ phase space that implies higher spin-down luminosities. The orange dashed-dotted line corresponds to a single contour for magnetic field strength at the light cylinder, $B_{\rm{LC}}=10^5$ G, which is often given as a threshold value for GP-emitting pulsars\cite{Kuzmin2007}. A sample of giant pulse-emitting pulsars are indicated by orange diamonds\cite{Mickaliger2012,Bilous2015,Mahajan2018,McKee2019,Geyer2021}, which all reside above the $B_{\rm{LC}}$ threshold line.}
\label{fig:pulsar_compare}
\end{figure}

\section*{Comparison to radio magnetar observations}

Radio-loud magnetars have long been suspected as progenitors for FRB emission due to the energy requirements set by FRB luminosities and the burst properties that are qualitatively similar (i.e., temporal durations, spectral features, etc.)\cite{Metzger2017}. These suspicions were reinforced by the observations of bright radio bursts from a Galactic magnetar, SGR J1935+2154, with one of these bursts, FRB 20200428, approaching the luminosity of extragalactic FRBs\cite{Bochenek2020,chime/frb2020}. Followup observations at high sensitivity with FAST have demonstrated that regular pulses from the source appear over a narrow phase window out of phase with the x-ray pulsation profile, contrary to high luminosity bursts that appear in random phases\cite{Zhu2023}. This dichotomy suggests that unlike radio pulses, highly luminous bursts occur in random locations, triggered by explosive events in a dynamically evolving magnetosphere. 

Radio magnetars exhibit a number of properties that are considered anomalous compared to ordinary radio pulsars, including a flatter radio spectrum and greater temporal variability in radio flux, pulse-profile shape and polarisation fraction\cite{Kaspi2017}. The observed PA swing reported here for FRB 20221022A has been replicated in some magnetars (time integrated) profiles such as the radio-loud magnetar 1E 1547.0$-$5408, which displays PA evolution that suggests that the rotation and magnetic axes are nearly aligned, and a small beaming fraction\cite{Camilo2008, Lower2023}. Conversely, for another magnetar, XTE J1810$-$197, RVM fitting indicates a substantial misalignment between the rotation and magnetic axes, and a correspondingly large beaming fraction\cite{Camilo2007}. The polarisation properties reported here for FRB 20221022A are thus not inconsistent with existing observations of Galactic radio magnetars. For this reason we cannot rule out a magnetar progenitor for FRB 20221022A.

\section*{Implications for repetition rate/burst energy distribution} 

It is not currently known if repeating FRB sources are fundamentally different from those that have not been seen to repeat. Consideration of burst rate measurements alone cannot rule out the possibility that one-off FRBs arise from the same population as repeating sources\cite{chime/frb2023}. Within this framework, all FRB sources are predicted to repeat if observed for long enough and the observed morphological differences between the two samples\cite{Pleunis2021} do not necessarily reflect a fundamental difference in the central engines of the two samples. Indeed, the longer burst durations regularly seen from repeating sources can naturally be explained as a selection effect introduced from beamed emission whereby FRB sources with larger opening angles would simultaneously be observed to have larger burst widths and greater likelihood of repeat detection\cite{Connor2020}. In such a scenario, a positive correlation would be observed between repetition rate, beaming angle and pulse duration. 

The PA swing reported here from FRB 20221022A strongly disfavours cataclysmic models for the emission and suggests the possibility of detecting repeat bursts from the source. Given the high level of exposure of CHIME to FRB 20221022A (see Methods), the absence of detection of repeat bursts from this source implies a repetition rate of $\sim1/1090$ bursts hr$^{-1}$. Such a burst rate, although low, is not anomalously low relative to other CHIME/FRB one-off events at similar declinations\cite{chime/frb2021}. However, FRB 20221022A is significantly brighter than the likely fluence completeness ($95\%$) threshold of CHIME at this sky position.

\begin{figure}
\centering
\includegraphics[width=0.9\textwidth]{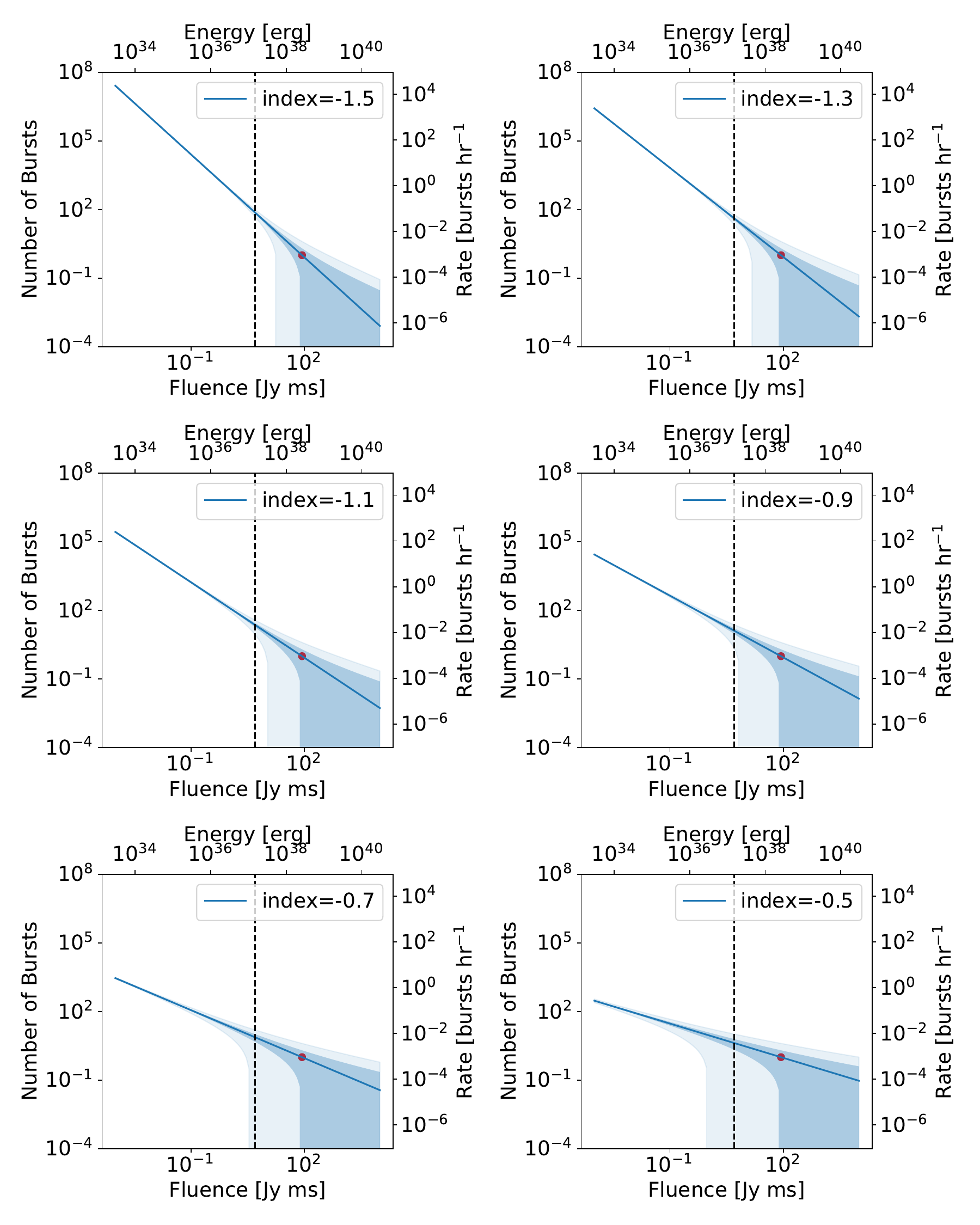}
\caption{The cumulative energy distribution representing number of burst detections above a certain fluence for a sample of power-law indices ($-1.5,-1.3,-1.1,-0.9,-0.7,-0.5$). The red point represents a single burst detected from FRB 20221022A. Burst rates represented as a twin vertical axis have been calculated from only the upper transit exposure (1090 hours) where the burst was detected. Blue shaded regions indicate 1 \& 3 $\sigma$ uncertainties determined assuming Poisson counting errors ($\sigma=\sqrt{\rm{N_{bursts}}}$). The vertical dashed line at 5 Jy ms is representative of a typical fluence threshold for a random source detected by CHIME/FRB. For a steep power-law (i.e., $|\gamma|>1$), the number of expected burst detections with fluence$\gtrsim 5$ Jy ms significantly exceeds 1.}
\label{fig:power_law_E}
\end{figure}

The absence of repeat detections of any fainter bursts thus can be used to put constraints on the burst energy distribution of the source. If we assume a power-law scaling for cumulative burst energy distribution, $N_{\rm{bursts}}(E>E_0) \propto E^{\gamma}$, we can constrain $\gamma$ by the absence of fainter repeat bursts from FRB 20221022A. Extended Data Fig.~\ref{fig:power_law_E} summarizes this analysis for different trial indices of a power-law cumulative energy distribution where $\gamma=-1.5$ is clearly inconsistent with observations, predicting $\sim 100$ bursts above a representative fluence completeness threshold of 5 Jy ms over the exposure duration. The absence of any detection of fainter bursts from FRB 20221022A therefore implies a relatively flat burst energy distribution. The constraint on the $\gamma$ of FRB 20221022A is summarized in Extended Data Fig.~\ref{fig:energy_index} and compared to equivalent measurements for a sample of repeating FRB sources\cite{Kirsten2023, Zhang2023, Nimmo2023, Zhang2023b}, giant pulse-emitting pulsars\cite{Mickaliger2012,Bilous2015,Mahajan2018,McKee2019,Geyer2021}, and Galactic magnetars\cite{Caleb2022,Wang2023b}. Our observations generally imply a much flatter burst energy distribution for FRB 20221022A than for the other sources. However, repeating source FRB 20201124A and to a lesser extent FRB 20200120E display emission modes wherein much flatter energy distributions are observed. In the case of FRB 20201124A, this flattening appears at spectral burst energies of $E_{\nu}\gtrsim 10^{31}$ erg Hz$^{-1}$ and is suspected to be a property of the high-energy segment of the burst energy distribution. This is significantly larger than the equivalent measurement for FRB 20221022A, $E_{\nu}\sim 6.5 \times 10^{29}$ erg Hz$^{-1}$, suggesting that if FRB 20221022A is repeating, the flattening of its burst energy distribution occurs at much lower energies than for FRB 20201124A.  

\begin{figure}
\centering
\includegraphics[width=0.99\textwidth]{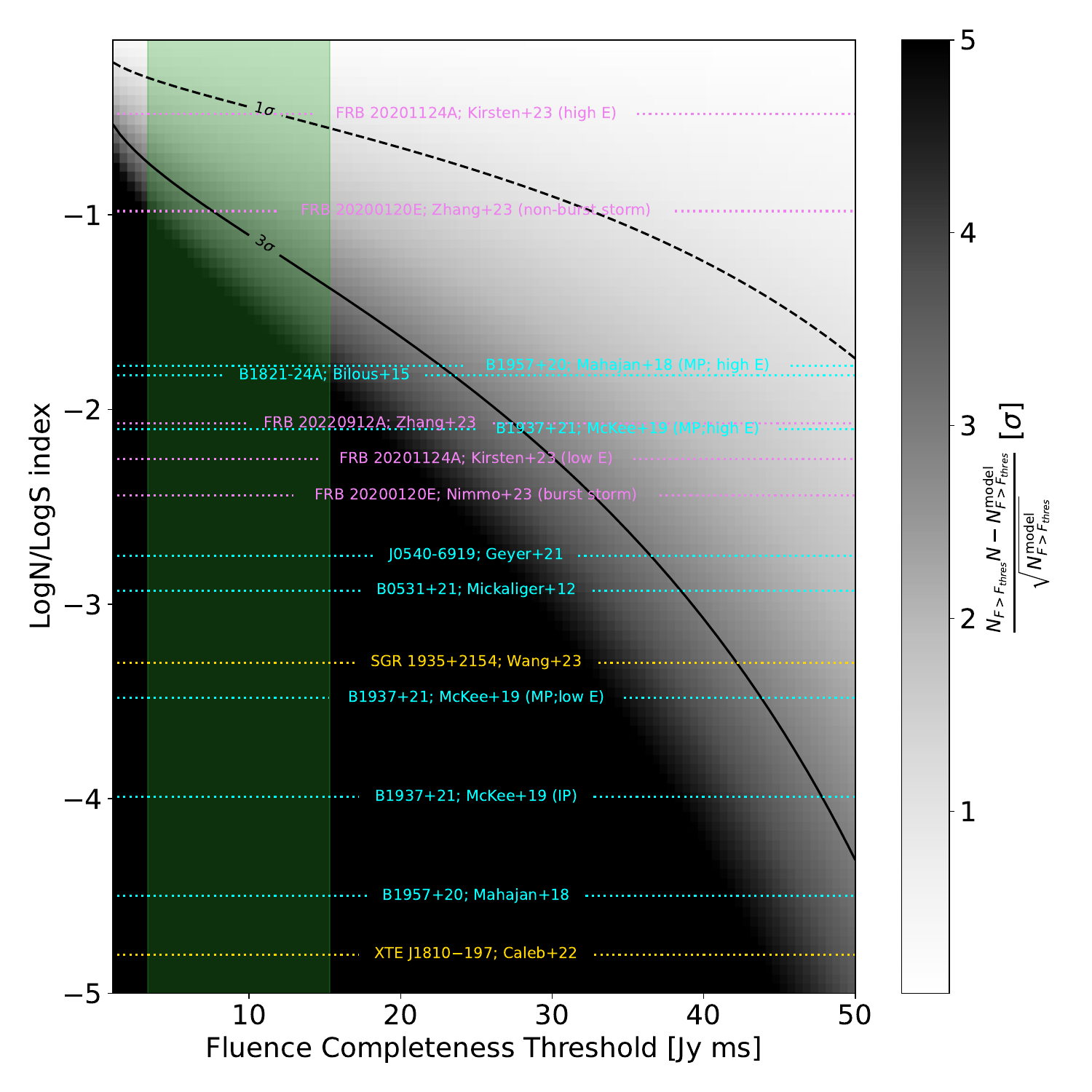}
\caption{Constraints on the power-law cumulative burst luminosity index ($\gamma$) for FRB 20221022A under different assumptions for the fluence ($95 \%$) completeness threshold for the source's transit in the main lobe of CHIME's primary beam. Darker regions of the grid indicate greater disagreement between the observed number of detections above the fluence threshold, $N_{F>\rm{thres}}=1$, and model predictions, $N^{\rm{model}}_{F>\rm{thres}}$, scaled by $\sqrt{N_{F>\rm{thres}}^{\rm{model}}}$ to factor counting error. Contour lines highlight regions of the parameter space that can be excluded at 1$\sigma$ \& 3$\sigma$ confidence. The vertical green band represents an estimate of the fluence completeness threshold for FRB 20221022A, determined from equivalent thresholds reported for other high ($\geq 85$ degrees) declination sources reported in ref.\cite{chime/frb2021}. Equivalent constraints on the power-law index for a sample of repeating FRB sources\cite{Kirsten2023, Zhang2023, Nimmo2023, Zhang2023b}, GP-emitting pulsars\cite{Mickaliger2012,Bilous2015,Mahajan2018,McKee2019,Geyer2021}, and Galactic magnetars\cite{Caleb2022,Wang2023b}, are indicated by violet, cyan and gold horizontal lines, respectively.}
\label{fig:energy_index}
\end{figure}

\section*{Source sensitivity \& exposure estimates} 
The CHIME/FRB exposures are estimated utilizing the methodology outlined in ref.\cite{chimefrbcatalog1}. In summary, the approach incorporates system uptime and sensitivity information spanning from August 2018 to December 2022, in conjunction with the beam model \footnote{\url{https://chime-frb-open-data.github.io/beam-model/}}, to generate healpix exposure maps. Subsequently, these healpix maps are queried at a specific source localization to derive exposure values. It is noteworthy that exposure calculations are exclusively performed for source transits occurring within the Full Width Half Maximum (FWHM) region of the beam at 600 MHz. Furthermore, sources situated above a declination of $\sim$70 degrees are circumpolar and transit the CHIME field of view twice within a sidereal day. We categorize them into upper and lower transits, relative to the zenith. Distinct exposure values are provided for these transits due to discernible variations in sensitivity between the two. 

FRB 20221022A is at a sufficiently high declination to be observable twice within a sidereal day by CHIME. In the aforementioned time range, CHIME/FRB has accumulated a total of 2752 hours of exposure on the source; 1090 hours in upper transit and 1662 hours in lower transit where the instrument is comparatively less sensitive. Sensitivity estimates are often reported as fluence completeness thresholds, indicating our confidence in detecting bursts brighter than this limit that occur within the exposure time. In the case of CHIME/FRB, fluence thresholds can differ substantially from source to source due to a number of factors including differences in the position of the sources within the static FFT search beams. For CHIME/FRB Catalog 1 sources\cite{chime/frb2021}, the fluence completeness thresholds of circumpolar sources have upper and lower transit fluence thresholds that roughly differ by factors $0.2-0.7$. For FRB 20221022A, rather than make a robust measurement of the fluence threshold, we arrive at a conservative estimate of the rate by only considering upper transit and using the $95\%$ fluence completeness thresholds reported in the CHIME/FRB Catalog\cite{chime/frb2021}. For catalog sources at similar or higher declinations to FRB 20221022A, we find that the 5th and 95th percentiles of the (upper-transit) fluence thesholds of this subsample are 3.3 and 13.3 Jy ms, respectively. This range of values is represented in Extended Data Fig.~\ref{fig:energy_index} as a green band where we find that even assuming a conservative upper limit on the fluence threshold for FRB 20221022A implies a cumulative burst energy distribution index that is inconsistent with benchmark value of $\gamma=-1.5$ that is often used in literature.

\section*{Search for repetition in CHIME/FRB data}
We check for association of known sources with every pulse detected by CHIME/FRB using a real-time actor called the 'Known Source Sifter`\cite{Pleunis2021b}. This burst was not associated with any other source in our known source database, including sources from ATNF, the Pulsar Survey scraper\cite{Kaplan2022}, and any FRB detected by CHIME/FRB. Given that this is a high declination event, potentially associated CHIME/FRB bursts would have a relatively unconstrained RA from hour angle information alone. Hence, we also run an in-house clustering algorithm developed for the identification of repeating FRB sources\cite{chime/frb2023} that effectively searches the CHIME/FRB database for any significant clusters across in RA, Dec and DM. For FRB 20221022A, we run the clustering algorithm on all CHIME detected events detected between 2018 July 25 and 2022 December 10 with a real-time (Bonsai) $S/N\geq 8.5$ and find no association with any other CHIME/FRB events.

\section*{Galactic DM contribution}
Given the low DM of this source, it is important to consider the possibility that the signal is coming from a source within the Galaxy, such as a pulsar. We explore this question using models of the free electron column, observations of pulsars with small angular separation, and low-DM FRBs. Galactic free electron models NE2001\cite{ne2001} and YMW16\cite{ymw17}, based on pulsars with independent DM and distance measurements, both predict maximum Galactic contributions along FRB\,20221022A's LoS of $\sim$60 pc cm$^{-3}$. Using a model of the scale height and midplane density of the warm ionized medium\cite{Ocker2020}, we estimate a Galactic disk contribution of $56\pm 5$ pc cm$^{-3}$. Of course, while the majority of pulsars sit within the Galactic disk, there is still the possibility of a source farther within the halo. Similarly, given the relatively low latitude, there is some possibility of increased DM contributions from low-latitude structures like HII regions. Estimates for the contribution of MW Halo DM span more than an order of magnitude, owing to the diffuse nature of ionized plasma. Observations from CHIME/FRB suggest upper-limits ranging between $52-111$ pc cm$^{-3}$ at higher latitudes, but support popular halo models and observational upper-limit estimates from ref.\cite{dgbb15,Yamasaki2020,Ravi2022}. The lower end of that range is an upper limit from the FRB repeater in M81, FRB\,20200120E\cite{Bhardwaj2021}. A lower latitude observational constraint of the MW halo DM, which similarly supports the extragalactic nature of FRB\,20221022A, is the recently detected one-off FRB 20220319D detected by the Deep Synoptic Array (DSA-110)\cite{Ravi2022}. This information, which supports but cannot unambiguously confirm the extragalactic nature of this source, is summarized in Extended Data Fig.~\ref{fig:DM}. The source has a DM larger than predicted by the Galactic electron density models, and the excess DM from the source is consistent with all considered models and upper-limits for the Milky Way's halo. 

\begin{figure}
    \centering
    \includegraphics[width=0.85\textwidth]{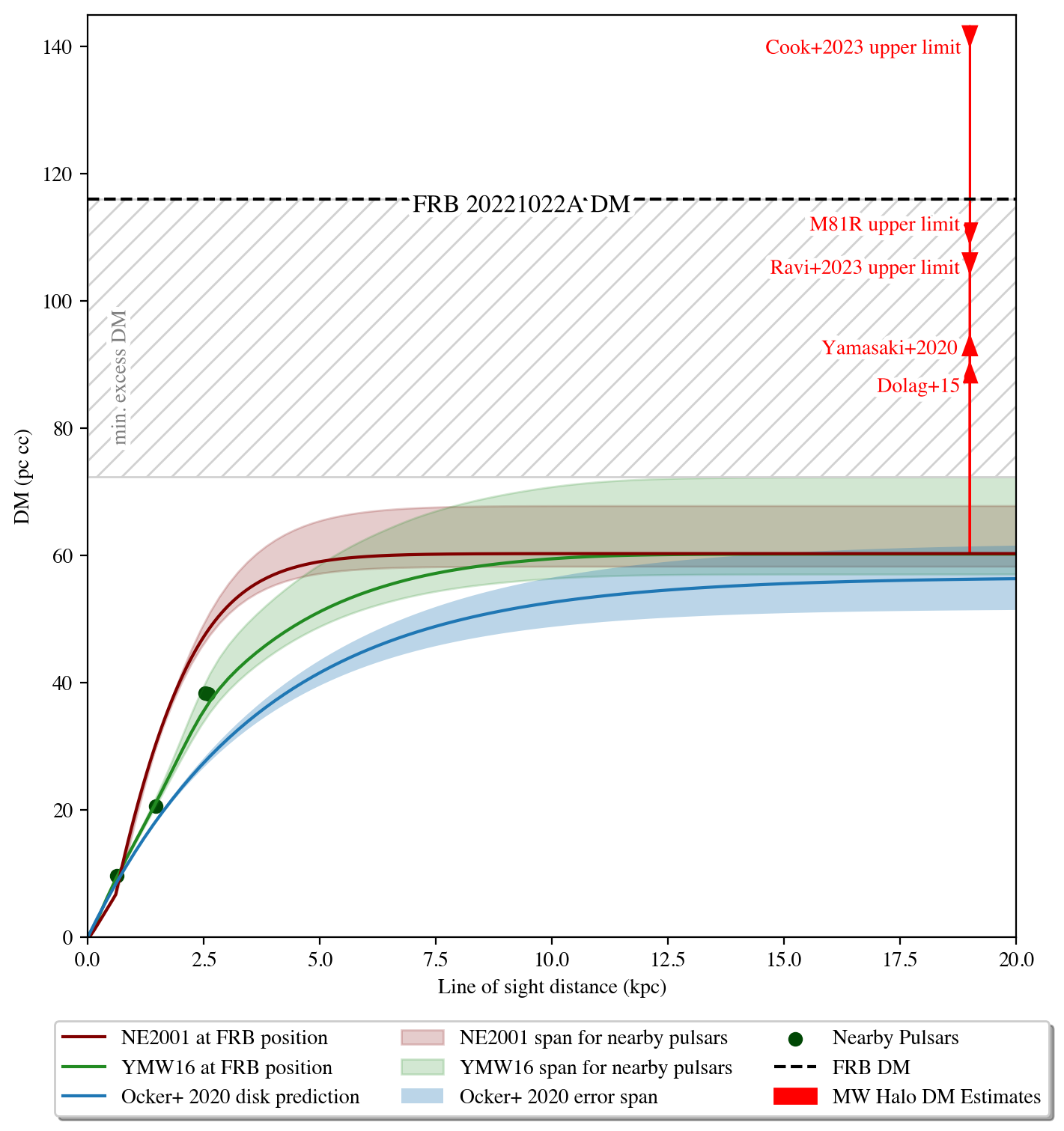}
    \caption{Dispersion measure versus LoS distance towards FRB 20221022A. The black dashed line shows the measured DM of FRB 20221022A. The estimate of Galactic DM contribution versus distance is shown from the YMW16 model (green line) and the mean value from the slab geometry model\cite{Ocker2020} (blue line). There is uncertainty associated with both of these estimates. For comparison, we search ATNF for pulsars with angular separation from the FRB less than 5 degrees, and plot the pulsars along with their DM and YMW16 predicted distance (from their measured DMs). We represent the spatial variation and uncertainty associated with the YMW16 Galactic free electron density model by showing the extent of the estimates along the lines of sight of each of the four `nearby' (on the sky) pulsars (green region). For the slab model, we show the extent of the one standard deviation errors (blue region). These estimates suggest a Galactic DM contribution of at least 78 pc cm$^{-3}$. These models do not include a contribution from the MW halo. Instead, we show estimates of the MW Halo contribution from a cosmological simulations\cite{Dolag2015, Prochaska2019}, a disk+sphere model based on X-ray emission measures\cite{Yamasaki2020}, and then two upper limits on DM Halo from observations of nearby FRBs, `Mark'\cite{Ravi2023} and FRB\,20200120E\cite{Bhardwaj2021}. Finally, we plot the largest observationally-supported Halo DM at $b = 30\deg$ from CHIME/FRB's first catalog\cite{Cook2023}, the closest line of sight included in the study.}
    \label{fig:DM}
\end{figure}

\end{methods}

%% file: acknowledgements.tex
\newcommand{\genacks}{
We acknowledge that CHIME is located on the traditional, ancestral, and unceded territory of the Syilx/Okanagan people.
We thank the Dominion Radio Astrophysical Observatory, operated by the National Research Council Canada, for gracious hospitality and expertise. 
CHIME is funded by a grant from the Canada Foundation for Innovation (CFI) 2012 Leading Edge Fund (Project 31170) and by contributions from the provinces of British Columbia, Qu\'ebec and Ontario. The CHIME/FRB Project is funded by a grant from the CFI 2015 Innovation Fund (Project 33213) and by contributions from the provinces of British Columbia and Qu\'ebec, and by the Dunlap Institute for Astronomy and Astrophysics at the University of Toronto. Additional support was provided by the Canadian Institute for Advanced Research (CIFAR), McGill University and the McGill Space Institute via the Trottier Family Foundation, and the University of British Columbia.
The Dunlap Institute is funded through an endowment established by the David Dunlap family and the University of Toronto. 
Research at Perimeter Institute is supported by the Government of Canada through Industry Canada and by the Province of Ontario through the Ministry of Research \& Innovation. 
The National Radio Astronomy Observatory is a facility of the National Science Foundation (NSF) operated under cooperative agreement by Associated Universities, Inc. 
FRB research at UBC is supported by an NSERC Discovery Grant and by the Canadian Institute for Advanced Research. 
The CHIME/FRB baseband system is funded in part by a Canada Foundation for Innovation John R. Evans Leaders Fund award to I.H.S.
This work has made use of data from the European Space Agency (ESA) mission {\it Gaia}(\url{https://www.cosmos.esa.int/gaia}), processed by the {\it Gaia} Data Processing and Analysis Consortium (DPAC,
\url{https://www.cosmos.esa.int/web/gaia/dpac/consortium}). Funding for the DPAC has been provided by national institutions, in particular the institutions participating in the {\it Gaia} Multilateral Agreement.
Observations presented here made use of the Gran Telescopio Canarias (GTC), installed at the Spanish Observatorio del Roque de los Muchachos of the Instituto de Astrof\'isica de Canarias, on the island of La Palma.

}